\documentclass[a4paper,11pt]{article}
\usepackage[margin=2cm]{geometry}
\usepackage{tikz,amsmath, amssymb,bm,color, amsthm}
\usetikzlibrary{positioning, calc,chains,fit,shapes}
\usepackage{enumerate}
\usepackage{comment}
\usepackage{tikz}
\usepackage{graphics}
\usepackage{longtable}
\usepackage{mdframed}
\usepackage{caption}
\usepackage{subcaption}
\usepackage{slashbox}
\usepackage{url}
\usepackage{framed}
\usepackage{array}
\usepackage{tabu}
\usepackage{lscape}
\usepackage{multirow}
\usepackage{ulem}
\usepackage{multicol}
\usepackage{placeins}
\usepackage{cite}
\usepackage{enumitem}

\usepackage{authblk}

\mdfsetup{skipabove=2pt,skipbelow=2pt}

\newtheorem{theorem}{Theorem}[section]
\newtheorem{lemma}[theorem]{Lemma}

\newtheorem{corollary}[theorem]{Corollary}
\newtheorem{proposition}[theorem]{Proposition}

\newtheorem{construction}{Construction}

\newcommand{\pname}[1]{\textnormal{\textsc{#1}}}

\newcommand{\FSAT}{\pname{4-SAT$_{\geq 2}$}}

\newcommand{\KSAT}{\pname{$k$-SAT$_{\geq {k-2}}$}}
\newcommand{\KSATO}{\pname{$k$-SAT}}

\newcommand{\FSATO}{\pname{4-SAT}}

\newcommand{\TSAT}{\pname{3-SAT}}

\newcounter{rowcntr}[table]
\renewcommand{\therowcntr}{\thetable.\arabic{rowcntr}}

\newcolumntype{N}{>{\refstepcounter{rowcntr}\therowcntr}c}

\AtBeginEnvironment{longtabu}{\setcounter{rowcntr}{0}}

\newcounter{rowcntra}[table]
\renewcommand{\therowcntra}{\arabic{rowcntra}}

\newcolumntype{M}{>{\refstepcounter{rowcntra}\therowcntra}c}

\AtBeginEnvironment{tabular}{\setcounter{rowcntra}{0}}

\newcommand{\SCG}{\pname{SC to $\mathcal{G}$}}
\newcommand{\SCGD}{\pname{SC to $\mathcal{G}'$}}

\newcommand{\SCT}{\pname{SC to}}
\newcommand{\SUBCOMTO}{\pname{Subgraph Complement to}}

\usetikzlibrary{fit}
\usetikzlibrary{
  graphs,
  graphs.standard
}
\author[1]{Dhanyamol Antony}
\author[2]{Jay Garchar}
\author[2]{Sagartanu Pal}
\author[2]{R. B. Sandeep}
\author[2]{Sagnik Sen}
\author[1]{R Subashini}

\affil[1]{National Institute of Technology Calicut, India

\texttt{\{dhanyamol\_p170019cs,suba\}@nitc.ac.in}}
\affil[2]{Indian Institute of Technology Dharwad, India

\texttt{\{170010001,183061001,sandeeprb,sen\}@iitdh.ac.in}}
\title{On subgraph complementation to $H$-free graphs}

\date{}
\begin{document}
\maketitle
\begin{abstract}
\end{abstract}

For a class $\mathcal{G}$ of graphs, the problem \textsc{Subgraph Complement to} $\mathcal{G}$ asks whether one can find a subset $S$ of vertices of the input graph $G$ such that 
complementing the subgraph induced by $S$ in $G$ results in a graph in $\mathcal{G}$.
We investigate the complexity of the problem when $\mathcal{G}$ is $H$-free for $H$
being a complete graph, a star, a path, or a cycle. We obtain the following results:
\begin{itemize}[leftmargin=*]
    \item When $H$ is a $K_t$ (a complete graph on $t$ vertices) for any fixed $t\geq 1$, the problem is solvable in polynomial-time.
    This applies even when $\mathcal{G}$ is a subclass of $K_t$-free graphs recognizable in polynomial-time, for example, the class of $(t-2)$-degenerate graphs.
    \item When $H$ is a $K_{1,t}$ (a star graph on $t+1$ vertices), we obtain that the problem is NP-complete for every $t\geq 5$.
    This, along with known results, leaves only two unresolved cases - $K_{1,3}$ and $K_{1,4}$. 
    \item When $H$ is a $P_t$ (a path on $t$ vertices), we obtain that the problem is NP-complete for every $t\geq 7$, leaving
    behind only two unresolved cases - $P_5$ and $P_6$.
    \item When $H$ is a $C_t$ (a cycle on $t$ vertices), we obtain that the problem is NP-complete for every $t\geq 8$, leaving 
    behind four unresolved cases - $C_4, C_5, C_6,$ and $C_7$.
\end{itemize}
Further, we prove that these hard problems do not admit 
subexponential-time algorithms (algorithms running in time $2^{o(|V(G)|)}$), assuming the Exponential Time Hypothesis.
A simple complementation argument implies that results for $\mathcal{G}$ are applicable for $\overline{\mathcal{G}}$,
thereby obtaining similar results for $H$ being the complement of a complete graph, a star, a path, or a cycle.
Our results generalize two main results and resolve one open question by Fomin et al. (Algorithmica, 2020).
\section{Introduction}
\label{sec:intro}

For a class $\mathcal{G}$ of graphs, a general graph modification problem can be defined as follows: 
Given a graph $G$, is there a set of modifications applying which on $G$ results in a graph in $\mathcal{G}$? 
Based on the type and the number of modifications allowed, there are various kinds of graph modification problems.
Among them, the most studied problems are vertex deletion problems and edge modification problems. As the names suggest,
the allowed modifications in them are vertex deletions and edge modifications (deletion, completion, or editing) respectively. 
In these types of graph modification problems, the size of the set of modifications is bounded by an additional integer input.
For example, in the \textsc{Cluster Editing} problem, given a graph $G$ and an integer $k$, the task is to find
whether there exists a set of at most $k$ pairs of vertices of $G$ such that changing the adjacencies of the pairs in $G$
results in a cluster graph.

In this paper, we deal with a graph modification known as subgraph complementation. In
subgraph complementation problems, the objective is to check whether the given graph $G$ has a subset $S$ of vertices
such that complementing the subgraph induced by $S$ in $G$, 
results in a graph in $\mathcal{G}$. 
Here, the adjacency of a pair $u,v$ of vertices is flipped only if both $u$ and $v$ are in $S$.
The graph
thus obtained, denoted by $G\oplus S$, is known as a subgraph complement of $G$. 
Unlike the vertex/edge modification problems, the operation is allowed only once but
there is no restriction on the size of $S$. 
This operation is introduced by Kami\'{n}ski et al.~\cite{DBLP:journals/dam/KaminskiLM09} as an attempt to generalize different kinds of complementations such 
as graph complementation and local complementation (replacing the closed neighborhood of a vertex by its complement) 
in their study of the clique-width of a graph. Recently, a systematic algorithmic study of this problem has been started by
Fomin et al.~\cite{DBLP:journals/algorithmica/FominGST20}.
They proved that the problem is polynomial-time solvable when the graph class $\mathcal{G}$
is triangle-free graphs, or $\mathcal{G}$ is $d$-degenerate, 
or $\mathcal{G}$ is of bounded clique-width and expressible in MSO\textsubscript{1} (for example, $P_4$-free graphs), 
or when $\mathcal{G}$ is the class of split graphs. They also obtained a hardness result in which they proved that
the problem is NP-complete if $\mathcal{G}$ is the class of regular graphs.

We focus on subgraph complementation problems for $\mathcal{G}$ being $H$-free graphs. 
For a graph $H$, in the problem \SUBCOMTO\ $H$-free graphs (\SCT\ $H$-free graphs), the task is to find whether there exists a subset $S$ of vertices of the input 
graph $G$ such that $G\oplus S$ is $H$-free, i.e., $G\oplus S$ does not contain any induced subgraph isomorphic to $H$.
There are numerous algorithmic studies on graph modification problems 
where the target graph class $\mathcal{G}$ is $H$-free~\cite{AravindSS17,CaiC15incompressibility,cai2012polynomial,Guo09more,CaoC12,GuillemotHPP13,CaoKY20,EibenLS15,DBLP:journals/siamcomp/Yannakakis81,DBLP:journals/jgt/JelinkovaK14}. We add on to this list by studying $H$-free graphs with respect to subgraph complementation.
A class $\mathcal{G}$ of graphs is hereditary (on induced subgraphs) if for every $G\in \mathcal{G}$, 
every induced subgraph of $G$ is in $\mathcal{G}$. It is well known that every hereditary class of graphs can be 
characterized by a set $\mathcal{H}$ of forbidden induced subgraphs. Therefore, studying a graph modification problem with target graph class
$H$-free (i.e., $|\mathcal{H}|=1$) can be seen as a first step toward understanding the complexity of the problem for hereditary properties.

We consider four classes of graphs $H$ - complete graphs, stars, paths, and cycles - and their complement classes. 
In all these cases, we obtain complete polynomial-time/NP-complete dichotomies, except for a few cases.
The results are summarized below:
\begin{itemize}[leftmargin=*]
    \item When $H$ is a $K_t$ (a complete graph on $t$ vertices), for any fixed $t\geq 1$, we obtain that \SCT\ $H$-free graphs can be solved in polynomial-time.
We obtain this result by generalizing the technique used in~\cite{DBLP:journals/algorithmica/FominGST20} for \SCT\ triangle-free graphs
and using results on generalized split graphs from~\cite{DBLP:conf/fsttcs/KolayP15}.
Our result applies to any subclass of $K_t$-free graphs recognizable in polynomial-time.
This result, as far as the existence of polynomial-time algorithms is concerned, subsumes  
two main results in~\cite{DBLP:journals/algorithmica/FominGST20} - the results when $\mathcal{G}$ is triangle-free
and when $\mathcal{G}$ is $d$-degenerate ($d$-degenerate graphs are $K_{d+2}$-free graphs).

\item When $H$ is a $K_{1,t}$ (a star graph on $t+1$ vertices), we obtain that \SCT\ $H$-free graphs is NP-complete for every fixed $t\geq 5$.
When $t=1$ (i.e., $H=K_2$), the problem can be solved trivially - a graph $G$ is a yes-instance of \SCT\ $K_2$-free graphs
if and only if $G$ is a $K_s\cup tK_1$ (disjoint union of a clique and isolated vertices), which can be recognized in polynomial-time. 
When $t=2$ (i.e., $H=P_3$), the problem admits a polynomial-time algorithm as the class of $P_3$-free graphs has
bounded clique-width and can be expressed in MSO\textsubscript{1} -- see Section 6 in~\cite{DBLP:journals/algorithmica/FominGST20}.
Therefore, the only remaining cases to be solved among the star graphs are $K_{1,3}$ and $K_{1,4}$.

\item When $H$ is a $P_t$ (a path on $t$ vertices), 
we obtain that \SCT\ $H$-free graphs is NP-complete for every fixed $t\geq 7$.
It is known from~\cite{DBLP:journals/algorithmica/FominGST20} that the problem can be solved in polynomial-time for all $t\leq 4$.
Therefore, the only remaining cases to be solved here are $P_5$ and $P_6$.

\item When $H$ is a $C_t$ (a cycle on $t$ vertices), we obtain that \SCT\ $H$-free graphs is NP-complete for every $t\geq 8$.
Therefore, the only remaining unknown cases among cycles are $C_4, C_5, C_6,$ and $C_7$.

\item We prove that a graph $G$ is a yes-instance of \SCT\ $\mathcal{G}$
if and only if $\overline{G}$ is a yes-instane of \SCT\ $\overline{\mathcal{G}}$, where $\overline{\mathcal{G}}$ is the set of 
complements of graphs in $\mathcal{G}$. 
This implies that \SCT\ $\mathcal{G}$ can be solved in polynomial-time if and only if \SCT\ $\overline{\mathcal{G}}$ can be solved in polynomial-time.
This resolves an open question in~\cite{DBLP:journals/algorithmica/FominGST20}.
Further, it implies that \SCT\ $H$-free graphs is polynomially equivalent to \SCT\ $\overline{H}$-free graphs.
Therefore, all our results for $H$-free graphs are applicable for $\overline{H}$-free graphs as well.

\item We observe that \SCT\ $\mathcal{G}$, for any polynomial-time recognizable class $\mathcal{G}$ of graphs, can be solved in time $2^{O(|V(G)|})$
by checking whether every subset $S$ of vertices in the input graph $G$ is a solution or not.
We obtain that this is the best one can hope, for all the cases in which we prove the NP-completeness. 
To be precise, we prove that, assuming the Exponential Time Hypothesis, 
there exists no subexponential-time algorithm (algorithm running in time $2^{o(|V(G)|)}$)
for every problem for which we prove the NP-completeness.

\end{itemize}

For the hardness results, we employ two types of reductions. One type, from variants of \textsc{SAT} problems, is used to solve base cases - 
for example $K_{1,5}$. The other type of reductions acts as an inductive step - for example from \SCT\ $K_{1,t}$-free 
graphs to \SCT\ $K_{1,{t+1}}$-free graphs. This scheme of obtaining hardness results was used to obtain a complete polynomial-time/NP-complete
dichotomy in \cite{AravindSS17}, and a conditional kernelization complexity dichotomy in \cite{DBLP:conf/esa/MarxS20} for $H$-free edge modification problems.
We believe that our results and techniques will come in handy for an eventual complexity dichotomy for \SCT\ $H$-free graphs.  

The paper is organized as follows: Preliminaries are given in Section~\ref{sec:prelim}, 
structural results are obtained in Section~\ref{sec:structural}, polynomial-time
algorithms are discussed in Section~\ref{sec:poly}, and the hardness results are proved in Section~\ref{sec:hardness}.
\section{Preliminaries}
\label{sec:prelim}

    A simple graph is a pair $G=(V,E)$, where $V$ is a set of vertices and $E\subseteq \binom{V}{2}$ is a set of edges.  
    For a graph $G$, we refer to its vertex set as $V(G)$ and its edge set as $E(G)$.  For a graph $G$ and a set $S\subseteq V(G)$, the \textit{induced subgraph} $G[S]$ is a graph whose vertex set is $S$ and whose edge set contains all the edges in $E$ that have both endpoints in $S$.
    For a vertex $v \in V(G)$,  the open neighborhood of $v$, denoted by $N(v)$, is the set of all the vertices adjacent to $v$, i.e., 
    $N(v):=\{w $ $|$ $vw \in E(G)\}$, and the closed neighborhood of $v$, denoted by $N[v]$, 
    is defined as $N(v) \cup \{v\}$. 
    The degree of a vertex $v$ is the size of its open neighborhood. 
    A vertex $v$ is a degree-$k$ if $v$ has degree $k$. 
    By $G-X$, we denote the graph obtained from $G$ by removing the vertices in $X$, i.e., $G-X = G[V(G)\setminus X]$. 
    A set $X$ of vertices in a graph $G$ is said to be a \textit{module} if every vertex in 
    $X$ has the same set of neighbours outside of $X$. 
     An \textit{empty graph}  is a graph without any edges and a \textit{null graph} is a graph without any vertices.   
     Let $H$ be any graph. Then a graph $G$ is called \textit{$H$-free}, if $G$ does not contain $H$ as an induced subgraph.
    In a graph $G$, two sets of vertices are said to be \textit{all-adjacent}, 
    if each vertex in one set is adjacent to every vertex in the other set. 
    Similarly, two sets of vertices are said to be \textit{nonadjacent}, 
    if there are no edges between them.
    A complete graph, an empty graph, a star, a cycle, and a path with $t$ vertices are denoted by 
    $K_t, tK_1, K_{1,t-1}, C_t,$ and $P_t$ respectively. 
    By $I_t$, we denote an independent set of size $t$.
    The center vertex of a star graph $K_{1,t}$ (for any $t\geq 2$), is the vertex having degree $t$. 
    For a class $\mathcal{G}$ of graphs, by $\overline{\mathcal{G}}$, we denote the class of complements of graphs in $\mathcal{G}$.
    A graph property $\Pi$ is \textit{nontrivial} if it is true for infinitely many graphs and false for infinitely many graphs. 
    The property is said to be \textit{trivial} otherwise.
    The \textit{disjoint union} of two graphs $G_1$ and $G_2$, denoted by $G_1\cup G_2$, is the graph $G$ such that
    $V(G) = V(G_1)\cup V(G_2)$ and $E(G) = E(G_1) \cup E(G_2)$. The disjoint union of $t$ copies of a graph $G$ is denoted by $tG$.
    The \textit{cross product} $H\times H'$ of two graphs $H$ and $H'$ is a graph $G$ such that the vertex set 
    $V(G)= V(H) \times V(H')$ and two vertices $(u,u' )$ and $(v,v' )$ are adjacent in 
    $G$ if and only if either $u = v$ and $u'$ is adjacent to $v'$ in $H'$, or $u' = v'$ and $u$ is adjacent to $v$ in $H$. A \textit{$k$-degenerate} graph is an undirected graph in which every subgraph has a vertex of degree at most $k$. For a graph $G$, the \textit{degeneracy} of   $G$ is the smallest value of $k$ for which it is $k$-degenerate.

    As an attempt to generalize split graphs, Gy\'{a}rf\'{a}s~\cite{DBLP:journals/jct/Gyarfas98} introduced the notion of $(p, q)$-split graphs.
    For positive integers $p, q$, a graph is a $(p, q)$-split graph if its vertices can be partitioned into two
    sets $P$ and $Q$ such that the clique number of $G[P]$ is at most $p$ and the independence number of 
    $G[Q]$ is at most $q$, i.e., $G[P]$ is $K_{p+1}$-free and $G[Q]$ is $(q+1)K_1$-free. 
    Clearly, a split graph is a $(1,1)$-split graph.
    Every such partition of the form $(P, Q)$ will be called a $(p,q)$-split partition. 
    For integers $p, q$, by $R(p,q)$ we denote the Ramsey number, i.e., $R(p,q)$ is the minimum 
    integer $n$ such that every graph $G$ with at least $n$ vertices has either a clique of size $p$
    or an independent set of size $q$.
    
    We say that boolean formula is a \KSATO\ formula if it is in conjunctive normal form (CNF) and every 
    clause contains exactly $k$ literals of distinct variables.
    We denote the $n$ variables and $m$ clauses of a \KSATO\ formula $\Phi$ by
    $\{X_1, \dots, X_n\}$, and $\{C_1,C_2,\dots,C_m\}$ respectively.
    For each variable $X_i$, 
    we denote the positve literal by $x_i$ and the negative literal by $\overline{x_i}$.
    Each clause $C_i$ is a 
    disjunction of exactly $k$ literals $\ell_{i,1}, \ell_{i,2},\dots\ell_{i,k}$, i.e., $C_i=\ell_{i,1}\lor\ell_{i,2}\lor\dots\lor\ell_{i,k}$. 
    The problem \KSAT\ is defined below.
    
\begin{mdframed}
  \textbf{\KSAT}: Given a boolean formula $\Phi$ with  $n$ variables and $m$ clauses in conjunctive normal form (CNF), where each clause contains exactly $k$ literals of distinct variables, find whether there exists a satisfying assignment for $\Phi$ with at least $k-2$ true literals per clause.\hfill
\end{mdframed}

The Exponential-Time Hypothesis (ETH) along with the Sparcification Lemma imply that \TSAT\ cannot be solved in time $2^{o(n+m)}$, 
where $n$ is the number of variables and $m$ is the number of clauses in the input formula. 
To show that a graph problem does not admit an algorithm running in time $2^{o(|V(G)|)}$ 
(where $G$ is the input graph), it is sufficient to give a polynomial-time reduction from \TSAT\ 
such that the resultant graph has only $O(n+m)$ vertices. 
We can show the same by a reduction from another graph problem 
(which does not admit a $2^{o(n)}$ algorithm, where $n$ is the number of vertices in the input graph), such that the 
resultant instance has only at most $O(n)$ vertices. Such reductions, where the blow-up in the input size (with respect to an appropriate measure -- 
the number of vertices in our case) is only linear, are known as linear reductions.
We refer to Chapter 14 of \cite{book:pa} for an exposition to these topics.
Since all the problems discussed in this paper are trivially in NP, we will not 
state the same explicitly while proving the NP-completeness of the problems.

\begin{proposition}[folklore]
    For every $k\geq3$, \KSAT\ is NP-Complete. Further, the problem cannot be solved in time $2^{o(n+m)}$, assuming the ETH.
\end{proposition}

\begin{proof}
We prove by induction on $k$. When $k=3$, \KSAT\ is same as \TSAT. Assume that the statement is true
for some $k=s\geq 3$. We will prove that the statement is true for $k=s+1$ by showing a linear reduction
from \pname{$s$-SAT$_{s-2}$} to \pname{$(s+1)$-SAT$_{s-1}$}.
Let $\Phi$ be an instance of \pname{$s$-SAT$_{s-2}$} formula with $n$ variables $X_1,\dots, X_n$, 
and $m$ clauses $C_1,\dots , C_m$. 
We transform $\Phi$ into a \pname{$(s+1)$-SAT$_{s-1}$} instance $\Psi$ in the following way: 
for each $C_i$ where $C_i = \ell_{i,1}\lor \ell_{i,2} \lor \ldots \ell_{i,s}$, 
 we construct a clause $C'_i = \ell_{i,1}\lor \ell_{i,2} \lor \ldots \ell_{i,s}\lor y_i$, where
 $y_i$ is the positive literal of a new variable $Y_i$.
 Let the resultant formula be $\Psi$ and let  $Y=\{Y_1, Y_2, \ldots, Y_m\}$.
 Clearly, $\Psi$ has $n+m$ variables and $m$ clauses and hence the reduction is linear.
 Now, it is sufficient to prove that 
 $\Phi$ is satisfiable (i.e., there exists a truth assignment such that every clause has at least $s-2$ true literals) 
 if and only if $\Psi$ is satisfiable (i.e., there exists a truth assignment such that every clause has at least $s-1$ true literals). 
 
 To prove the forward direction, assume that $\Phi$ is satisfiable with a set $L$ of true variables.
 Since every clause of $\Phi$ has at least $s-2$ true literals, every clause of $\Psi$ has at least $s-1$ true literals if we 
 assign True to every variable in $Y$. Therefore, $L\cup Y$ satisfies $\Psi$. 
 
To prove the other direction, assume that $\Psi$ is satisfiable. That means, there is a truth assignment in which 
every clause has at least $s-1$ true literals. Therefore, by the same truth assignment (confined to the variables of $\Phi$), 
every clause in $\Phi$ has at least $s-2$ true literals. 
\end{proof}

    A \textit{subgraph complement} of a graph $G$ is a graph $G'$ obtained from $G$, for any $S\subseteq V(G)$, by complementing the subgraph induced by $S$ in $G$.
    More formally, $V(G') = V(G)$ and two vertices $u,v$ are adjacent in $G'$ if and only if at least one of the following conditions hold true - (i) $u$ and $v$
    are adjacent in $G$ and either $u$ or $v$ is not in $S$; (ii) $u$ and $v$ are nonadjacent in $G$ and both $u$ and $v$ are in $S$.

\begin{mdframed}
  \textbf{\SUBCOMTO\ $\mathcal{G}$ (\SCG)}: Given a graph $G$, is there a subgraph complement $G'$ of $G$ such that $G'\in \mathcal{G}$?\hfill
\end{mdframed}

By $\mathcal{G}^{(1)}$, we denote the class of graphs where each graph in it can be subgraph complemented to a graph in $\mathcal{G}$, i.e.,
$\mathcal{G}^{(1)} = \{G: \exists S\subseteq V(G)\ \text{such that}\ G\oplus S\in \mathcal{G}\}$. 

\section{Structural results}
\label{sec:structural}

For a graph $H$, let $\Pi_H$ be the property defined as follows: A graph $G$ has property $\Pi_H$ if there exists a set $S\subseteq V(G)$ such that $G\oplus S$ is $H$-free. That is, 
the class of graphs satisfying $\Pi_H$ is the set of all yes-instances of \SCT\ $H$-free graphs. We prove that $\Pi_H$ is nontrivial if and only if $H$ has at least two vertices. 
This result guarantees that the pursuit of obtaining optimal complexities of \SCT\ $H$-free graphs is meaningful for all nontrivial graphs $H$. 

\begin{lemma}
 \label{lem:nontrivial}
 For a graph $H$, the property $\Pi_H$ is nontrivial if and only if $H$ has at least two vertices.
\end{lemma}
\begin{proof}
  If $H$ has only one vertex, then none of the graphs with at least one vertex is a yes-instance of \SCT\ $H$-free graphs.
  Therefore, $\Pi_H$ is trivial. Now, let $H$ has at least two vertices. If $H$ has at least one edge, then every empty graph is a yes-instance of \SCT\ $H$-free graphs.
  If $H$ has no edges, then every complete graph is a yes-instance of \SCT\ to $H$-free graphs. 
  Now, it is sufficient to prove that there exists infinite number of no-instances of \SCT\ to $H$-free graphs.
  We claim that $G = \overline{H}\times H$ is a no-instance of \SCT\ to $H$-free graphs. 
  For a contradiction, assume that there exists a set $S\subseteq V(G)$ such that $G\oplus S$ is $H$-free.
  Since, the set formed by taking exactly one vertex from each copy of $\overline{H}$ induces an $H$, we obtain that every vertex of at least two
  copies of $\overline{H}$ is in $S$. Then there is a copy of $H$ in $G\oplus S$ induced by the set of vertices of a copy of $\overline{H}$ in $G$, which is a contradiction. 
\end{proof}



How is a subgraph complement of a graph $G$ related to a subgraph complement of $\overline{G}$ with respect to the same subset $S$ of vertices of the graphs? Lemma~\ref{lem:GS}
answers this question.
\begin{lemma}
\label{lem:GS}
   Let $G$ be a graph and $S\subseteq V(G)$. Then $G\oplus S$ = $\overline{\overline{G}\oplus S }$.
\end{lemma} 
\begin{proof}
  The statement is trivially true when $G$ has only one vertex. 
  Therefore, assume that $G$ has at least two vertices. Let $u,v$ be any two vertices in $G$.
  It is sufficient to prove that $u$ and $v$ are adjacent in $G\oplus S$ if and only if they are adjacent in $\overline{\overline{G}\oplus S}$.
  Assume that $u$ and $v$ are adjacent in $G\oplus S$. This gives rise to two cases: 
  
  Case 1: Both $u$ and $v$ are in $S$.  
  Then we obtain that $u$ and $v$ are nonadjacent in $G$, adjacent in $\overline{G}$, nonadjacent in $\overline{G}\oplus S$, and 
  adjacent in $\overline{\overline{G}\oplus S}$.
  
  Case 2: Either $u$ or $v$ is not in $S$.
  Then we obtain that $u$ and $v$ are adjacent in $G$, nonadjacent in $\overline{G}$, nonadjacent in $\overline{G}\oplus S$, and
  adjacent in $\overline{\overline{G}\oplus S}$.
  
  The case when $u$ and $v$ are nonadjacent in $G\oplus S$ can be proved analogously.  
\end{proof}

\begin{lemma}
\label{lem:G-G-complement}
For a class $\mathcal{G}$ of graphs, $\overline{\mathcal{G}^{(1)}} = \overline{\mathcal{G}}^{(1)}$.
\end{lemma}

\begin{proof}
 Let $G$ be a graph such that $G \in \overline{\mathcal{G}}^{(1)}$. 
 So there exists a set $S\subseteq V(G)$ such that $G\oplus S\in \overline{\mathcal{G}}$.
 Then by Lemma~\ref{lem:GS}, $\overline{\overline{G}\oplus S }\in \overline{\mathcal{G}}$. 
 Therefore, $\overline{G}\oplus S \in \mathcal{G}$. 
 This implies that $\overline{G}\in \mathcal{G}^{(1)}$ and hence
 $G \in \overline{\mathcal{G}^{(1)}}$. 
 Therefore, $\overline{\mathcal{G}}^{(1)}\subseteq\overline{\mathcal{G}^{(1)}}$.
 Now, it is sufficient to prove that $\overline{\mathcal{G}^{(1)}}\subseteq\overline{\mathcal{G}}^{(1)}$.
 Let  $G \in \overline{\mathcal{G}^{(1)}}$. 
 Then $\overline{G}\in \mathcal{G}^{(1)}$. That means there exist a set $S\subseteq V(\overline{G})$ 
 such that $\overline{G}\oplus S \in \mathcal{G}$. 
 Then $\overline{\overline{G}\oplus S }\in \overline{\mathcal{G}}$. 
 Then by Lemma~\ref{lem:GS}, $G\oplus S\in \overline{\mathcal{G}}$. 
 This implies $G\in \overline{\mathcal{G}}^{(1)}$. 
 Therefore, $\overline{\mathcal{G}^{(1)}}\subseteq\overline{\mathcal{G}}^{(1)}$. 
\end{proof}

Lemma~\ref{lem:G-G-complement} implies Corollary~\ref{cor:G-G-complement}, which tells us that \SCT\ $\mathcal{G}$ is polynomially equivalent to \SCT\ $\overline{\mathcal{G}}$.
The first statement in Corollary~\ref{cor:G-G-complement} was an open problem raised 
in~\cite{DBLP:journals/algorithmica/FominGST20}, whereas the second statement implies that all
results obtained in this paper for $H$-free graphs are applicable for $\overline{H}$-free graphs as well.

\begin{corollary}
\label{cor:G-G-complement}
For a class $\mathcal{G}$ of graphs, ${\mathcal{G}^{(1)}}$ can be recognized in polynomial time if and only if  $\overline{\mathcal{G}}^{(1)}$ can be recognized in polynomial time. 
In particular, \SCT\ $H$-free graphs is polynomial-time solvable if and only if \SCT\ $\overline{H}$-free graphs is polynomial-time solvable.
\end{corollary}

\section{Polynomial-time algorithms}
\label{sec:poly}

In this section, we obtain a polynomial-time algorithm for \SCG, 
when $\mathcal{G}$ is a subclass of $K_t$-free graphs, for any fixed integer $t\geq 1$, such that 
$\mathcal{G}$ is recognizable in polynomial-time.
To obtain this result, we generalize the technique used for \SCT\ triangle-free graphs in~\cite{DBLP:journals/algorithmica/FominGST20}.
Assume that a graph $G$ has a solution $S$, i.e., $G\oplus S\in \mathcal{G}$. 
Further, assume that $S$ has at least two vertices $u$ and $v$. 
Then, we prove that each of the sets $N(u)\cap N(v)$, $\overline{N[u]}\cap \overline{N[v]}$, $N(u)\setminus \overline{N[v]}$, and
$N(v)\setminus \overline{N[u]}$ induces a $(p,q)$-split graph (for $p = q = t-1$), and for each of them, $S$ contains exactly the set $Q$ of some $(p,q)$-split
partition $(P,Q)$ of the corresponding induced subgraph. 
Then the algorithm boils down to recognizing $(p,q)$-split graphs and enumerating all $(p,q)$-split partitions of them in 
polynomial-time. All the tools required for this task has already been obtained in~\cite{DBLP:conf/fsttcs/KolayP15}, 
see the full version \cite{kolay2015parameterized} for the proofs.

\begin{proposition}[\cite{DBLP:conf/fsttcs/KolayP15, kolay2015parameterized}]
\label{pro:split-diff}
Let $G$ be a $(p,q)$-split graph. Let $(P,Q)$ and $(P',Q')$ be two $(p,q)$-split partitions of $G$.
Then $|P\cap Q'|\leq R(p+1, q+1) - 1$ and $|P'\cap Q|\leq R(p+1, q+1) - 1$. 
\end{proposition}

Proposition~\ref{pro:split-diff} implies that two distinct $(p,q)$-partitions of a $(p,q)$-split graph
cannot differ too much. This helps us to enumerate all $(p,q)$-split partitions in polynomial-time, given one of them.
Proposition~\ref{pro:split-poly} says that there is a polynomial-time algorithm which recognizes a $(p,q)$-split graph and gives a 
$(p,q)$-split partition of the same.

\begin{proposition}[\cite{DBLP:conf/fsttcs/KolayP15, kolay2015parameterized}]
\label{pro:split-poly}
For any fixed constants $p$ and $q$, there is an algorithm which takes a graph $G$ (with $n$ vertices)
as an input, runs in time $O(n^{2R(p+1, q+1)+p+q+4})$, and decides whether $G$ is a $(p,q)$-split graph.
Furthermore, if $G$ is a $(p,q)$-split graph, then the algorithm outputs a $(p,q)$-split partition of $G$.
\end{proposition}
The running time in Proposition~\ref{pro:split-poly} is not explicitly given in \cite{DBLP:conf/fsttcs/KolayP15, kolay2015parameterized},
but can be easily derived from the proof given in \cite{kolay2015parameterized}.
The proof of the following lemma is very similar to that of Proposition~\ref{pro:split-poly}. 
\begin{lemma}
\label{lem:split-all}
Let $G$ be a $(p,q)$-split graph with $n$ vertices.
Then there are at most $n^{2R(p+1,q+1)}$ $(p,q)$-split partitions of $G$.
Given a $(p,q)$-split partition $(P,Q)$ of $G$,
all $(p,q)$-split partitions of $G$ can be computed in polynomial-time, specifically in time $O(n^{2R(p+1,q+1)+p+q+3})$.
\end{lemma}
\begin{proof}
  Let $(P',Q')$ be any $(p,q)$-split partition of $G$. By Proposition~\ref{pro:split-diff},
  $|P\cap Q'|\leq R(p+1, q+1)-1$ and $|P'\cap Q|\leq R(p+1, q+1)-1$.
  Let $X=P\cap Q'$ and $Y=P'\cap Q$. Clearly, there are at most $n^0 + n^1 +\ldots\ + n^{R(p+1, q+1)-1}\leq n^{R(p+1, q+1)}$ possible guesses for 
  $X$. Similarly, there are at most $n^{R(p+1, q+1)}$ guesses for $Y$. Therefore, there are at most 
  $n^{2R(p+1,q+1)}$ guesses for the pair $(X,Y)$. 
  Clearly, $P' = Y\cup (P\setminus X)$ and  $Q'=X\cup (Q\setminus Y)$.
  Hence, there are at most $n^{2R(p+1,q+1)}$ many $(p,q)$-split partitions of $G$.
  For each pair $(X,Y)$ of guesses, it is sufficient to check whether $(Y\cup (P\setminus X), X\cup (Q\setminus Y))$ is a 
  $(p,q)$-split partition of $G$. 
  For each pair $(X,Y)$ of guesses, choosing the pair takes linear time and the checking can be done in time $O(n^{p+q+2})$ - 
  it is sufficient to check that $(Y\cup (P\setminus X))$ does not induce a graph with a clique of size $p+1$ and $X\cup (Q\setminus Y)$
  does not induce a graph with an independent set of size $q+1$.
  Therefore, all $(p,q)$-split partitions of $G$ can be computed in time $O(n^{2R(p+1,q+1)+p+q+3})$. 
\end{proof}

Proposition~\ref{pro:split-poly} and Lemma~\ref{lem:split-all} directly imply Corollary~\ref{cor:split-all}.

\begin{corollary}
\label{cor:split-all}
For any fixed constants $p\geq 1$ and $q\geq 1$, there is an algorithm which takes a graph $G$ (with $n$ vertices)
as an input, runs in time $O(n^{2R(p+1, q+1)+p+q+4})$, and decides whether $G$ is a $(p,q)$-split graph.
Furthermore, if $G$ is a $(p,q)$-split graph, then the algorithm outputs all $(p,q)$-split partitions of $G$.
\end{corollary}

Let $G$ be a yes-instance of \SCT\ $\mathcal{G}$, where $\mathcal{G}$ is a subclass of $K_t$-free graphs recognizable 
in polynomial-time.
Assume that $G$ is not a trivial yes-instance. 
Let $S\subseteq V(G)$ be such that $G\oplus S\in \mathcal{G}$. 
Clearly, $|S|\geq 2$. Let $u,v$ be two vertices in $S$. 
With respect to $S, u, v$, we partition the vertices in $V(G)\setminus \{u,v\}$ into eight sets as given below.
This is depicted in Figure~\ref{kt}.

\begin{figure}[ht]
  \centering
    \centering
    \begin{tikzpicture}[myv/.style={ellipse, draw, inner xsep=50pt,inner ysep=25pt}, myv1/.style={ellipse, draw, inner xsep=80pt,inner ysep=40pt},myv2/.style={circle,color=white, draw,inner sep=0pt}, myv3/.style={ellipse, draw, inner sep=1.5pt},myv4/.style={circle, draw, inner sep=1.5pt}, myv5/.style={ellipse, draw, inner sep=1.5pt}]

\node[myv2] (a)[label= right:{$S_{\overline{uv}}$}] at (0.6,-0.8) {}; \node[myv2] (b)[label= right:{$S_{uv}$}] at (0.6,0.9) {}; 
\node[myv2] (c)[label= right:{$S_{{u}\overline{v}}$}] at (-1.4,0) {};
\node[myv2] (d)[label= right:{$S_{\overline{u}{v}}$}] at (2.5,0) {};
\node[myv2] (e)[label= right:{$T_{{u}\overline{v}}$}] at (-2.3,0) {};
\node[myv] (f) [label= right:{$T_{\overline{u}v}$}] at (1,0) {};
\node[myv1] (G)[label= right:{}] at (1,0) {};
\node[myv2] (h)[label= right:{$T_{uv}$}] at (0.6,1.6) {};
\node[myv2] (j)[label= right:{$T_{\overline{uv}}$}] at (0.6,-1.6) {};

  \node[myv4] (u) at (0.25,0) {$u$};
  \node[myv4] (v) at (1.75,0) {$v$};
 \node[myv5][fit= (u) (v),  inner xsep=0.25ex, inner ysep=0.25ex] {}; ] {};

 

  \draw (-0.2,1.9) -- (0.7,0.35);
  \draw (2.3,1.9) -- (1.3,0.35);
  \draw (0.7,-0.35) -- (-0.2,-1.9);
  \draw (1.3,-0.35) -- (2.3,-1.9);
  \draw (u)[line width=0.5mm] -- (0.9,0.8);
  \draw (u) [line width=0.5mm]-- (-2,1);
  \draw (u) [line width=0.5mm]-- (-0.5,0.8);
  \draw (u) [line width=0.5mm]-- (1,1.5);
  \draw (v)[line width=0.5mm] -- (0.9,0.8);
  \draw (v) [line width=0.5mm]-- (4,1);
  \draw (v) [line width=0.5mm]-- (2.5,0.8);
   \draw (v) [line width=0.5mm]-- (1,1.5);

\end{tikzpicture}
    \caption{Partitioning of vertices of a yes-instance $G$ of \SCT\ $\mathcal{G}$, 
    based on a solution $S$ and two vertices $u,v$ in $S$. 
    The bold lines represent the adjacency of  vertices $u$ and $v$. 
    } 
    \label{kt}
  \end{figure}
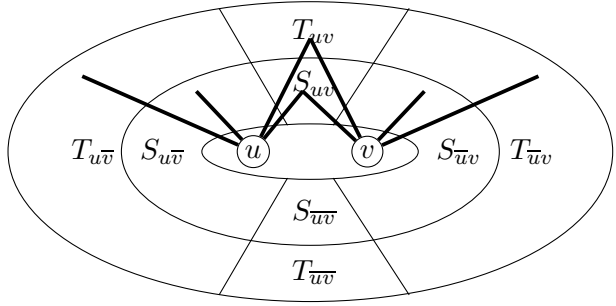
\begin{multicols}{2}
\begin{enumerate}[label=(\roman*)]
	\item $S_{uv}=S\cap N(u) \cap N(v)$
	\item $S_{\overline{uv}} = S\cap \overline {N[u]} \cap \overline{N[v]}$ 
	\item $S_{u\overline{v}}=S\cap (N(u)\setminus N[v])$ 
	\item $S_{\overline{u}v}=S\cap (N(v)\setminus N[u])$
	\item $T_{uv}=(N(u)\cap N(v))\setminus S$
	\item $T_{\overline{uv}}=(\overline{N[u]}\cap\overline{N[v]})\setminus S$
	\item $T_{u\overline{v}}=(N(u)\setminus N[v])\setminus S$
	\item $T_{\overline{u}v}=(N(v)\setminus N[u])\setminus S$
\end{enumerate}
\end{multicols}

Clearly, $S=S_{uv}\cup S_{\overline{uv}}\cup S_{u\overline{v}}\cup S_{\overline{u}v}\cup \{u,v\}$, 
and $V(G)\setminus S = T_{uv}\cup T_{\overline{uv}}\cup T_{u\overline{v}}\cup T_{\overline{u}v}$.

\begin{lemma}
\label{lem:kt-free-solution}
  Let $G$ be a yes-instance of \SCT\ $\mathcal{G}$, where $\mathcal{G}$ is a subclass of $K_t$-free graphs, for any fixed integer $t\geq 2$.
  Let $S\subseteq V(G)$ be such that $|S|\geq 2$ and $G\oplus S\in \mathcal{G}$. 
  Let $u$ and $v$ be any two vertices in $S$. Then the following statements hold true:
  \begin{enumerate}[label=(\roman*)]
      \item $N(u)\cap N(v)$ induces a $(t-1,t-1)$-split graph with a $(t-1,t-1)$-split partition $(T_{uv}, S_{uv})$;
      \item $\overline{N[u]}\cap \overline{N[v]}$ induces a $(t-1,t-1)$-split graph with a $(t-1,t-1)$-split partition $(T_{\overline{u}\overline{v}}, S_{\overline{u}\overline{v}})$;
      \item $N(u)\setminus N[v]$ induces a $(t-1,t-1)$-split graph with a $(t-1,t-1)$-split partition $(T_{u\overline{v}}, S_{u\overline{v}})$;
      \item $N(v)\cap N[u]$ induces a $(t-1,t-1)$-split graph with a $(t-1,t-1)$-split partition $(T_{\overline{u}v}, S_{\overline{u}v})$;
  \end{enumerate}
\end{lemma}
\begin{proof}
  To prove (i), we note that $S_{uv}\cup T_{uv} = N(u)\cap N(v)$.
  If $G[S_{uv}]$ has an independent set of size $t$, then $G\oplus S$ will have a clique of size $t$, which is a contradiction.
  Similarly, if $G[T_{uv}]$ has a clique of size $t$, then $G\oplus S$ will have a clique of size $t$, which is a contradiction.
  Therefore, $G[T_{uv}]$ is $K_t$-free and $G[S_{uv}]$ is $tK_1$-free.
  Hence, $N(u)\cap N(v) = T_{uv}\cup S_{uv}$ induces a $(t-1, t-1)$-split graph and $(T_{uv}, S_{uv})$ is a 
  $(t-1,t-1)$-split partition of it.
  The statements (ii), (iii), and (iv) can be proved in analogous ways.
\end{proof}
\begin{mdframed}
  \textbf{Algorithm for \SCT\ $\mathcal{G}$, where $\mathcal{G}$ is a subclass of $K_t$-free graphs}\\
  Input: A graph $G$\\
  Output: If $G$ is a yes-instance of \SCT\ $\mathcal{G}$, 
  then returns a set $S\subseteq V(G)$ such that $G\oplus S\in \mathcal{G}$; returns `None' otherwise.
  \begin{description}
      \item[Step 0]: If $G\in \mathcal{G}$, then return $\emptyset$.
      \item[Step 1]: For every unordered pair of vertices $\{u,v\}$ in $G$:
      \begin{enumerate}[label=(\roman*)]
          \item If any of 
          $N(u)\cap N(v)$, $\overline{N[u]}\cap \overline{N[v]}, N(u)\setminus N[v], N(v)\setminus N[u]$ 
          does not induce a $(t-1,t-1)$-split graph, then continue (with Step 1).
          \item Compute $L_{uv}$, the list of all $(p,q)$-split partitions of $N(u)\cap N(v)$.
          \item Compute $L_{\overline{u}\overline{v}}$, the list of all $(p,q)$-split partitions of $\overline{N[u]}\cap \overline{N(v)}$.
          \item Compute $L_{u\overline{v}}$, the list of all $(p,q)$-split partitions of $N(u)\setminus N[v]$.
          \item Compute $L_{\overline{u}v}$, the list of all $(p,q)$-split partitions of $N(v)\setminus N[v]$.
          \item For every $(P_a, Q_a)$ in $L_{uv}$, for every $(P_b, Q_b)$ in $L_{\overline{u}\overline{v}}$,
          for every $(P_c, Q_c)$ in $L_{u\overline{v}}$, and for every $(P_d, Q_d)$ in $L_{\overline{u}v}$:
          \begin{enumerate}
              \item Let $S=Q_a\cup Q_b\cup Q_c\cup Q_d\cup \{u,v\}$
              \item If $G\oplus S\in \mathcal{G}$, then return $S$.
          \end{enumerate}
      \end{enumerate}
      \item[Step 2]: Return `None'
  \end{description}
\end{mdframed}

\begin{theorem}
\label{thm:kt-free}
For any fixed $t\geq 1$, let $\mathcal{G}$ be a subclass of $K_t$-free graphs such that there is a recognition algorithm
for $\mathcal{G}$ running in time $O(f(n))$, for some polynomial function $f$. Then \SCT\ $\mathcal{G}$ can be solved in polynomial-time, specifically in time
$O(f(n)\cdot(n^{8R(t,t)+4}))$.
In particular, \SCT\ $K_t$-free graphs can be solved in polynomial-time, specifically in time  $O(n^{8R(t,t)+t+4})$.
\end{theorem}
\begin{proof}
  If $t=1$, then $G$ is a yes-instance if and only if $G$ is a null graph. Therefore, let $t\geq 2$.
  We claim that the algorithm given for \SCT\ $\mathcal{G}$ is correct. 
  The case when $G\in \mathcal{G}$ is handled correctly by Step 0.
  Therefore, assume that $G\notin \mathcal{G}$. 
  Assume that $G$ is a yes-instance, i.e., there exists a set $S$ such that
  $G\oplus S\in \mathcal{G}$. 
  Clearly $S$ contains at least two vertices, say $u'$ and $v'$.
  For a contradiction, assume that the algorithm returns `None'. 
  It means that the algorithm was unsuccessful in returning $S$ while processing the pair $(u=u', v=v')$.
  By Lemma~\ref{lem:kt-free-solution}, each of $N(u)\cap N(v)$, $\overline{N[u]}\cap \overline{N[v]}$, 
  $N(u)\setminus N[v]$, and $N(v)\setminus N[u]$ induces $(t-1,t-1)$-split graphs.
  Further, $(T_{uv}, S_{uv})$, $(T_{\overline{u}\overline{v}}, S_{\overline{u}\overline{v}})$, 
  $(T_{u\overline{v}}, S_{u\overline{v}})$, and $(T_{\overline{u}v}, S_{\overline{u}v})$
  are $(t-1,t-1)$-split partitions of the graphs induced by them. 
  Therefore, $(T_{uv}, S_{uv})\in L_{uv}$, 
  $(T_{\overline{u}\overline{v}}, S_{\overline{u}\overline{v}})\in L_{\overline{u}\overline{v}}$,
  $(T_{u\overline{v}}, S_{u\overline{v}})\in L_{u\overline{v}}$, and 
  $(T_{\overline{u}v}, S_{\overline{u}v})\in L_{\overline{u}v}$.
  Therefore, the algorithm correctly returns $S=S_{uv}\cup S_{\overline{u}\overline{v}}\cup S_{u\overline{v}}\cup S_{\overline{u}v}\cup \{u,v\}$
  in Step 1(vi)b, a contradiction to our assumption that the algorithm returns `None'.
  Whenever the algorithm returns a set $S$, clearly, $S$ is a solution as it is checked in Step 1(vi)b that $G\oplus S\in \mathcal{G}$. 
  Now, assume that $G$ is a no-instance. Then the algorithm correctly returns `None', as there exists no set $S$ such that $G\oplus S\in \mathcal{G}$ 
  and the condition in Step 1(vi)b always fails.
  
  To analyse the running time, assume that $G$ has $n$ vertices.
  Consider an iteration of Step 1, with a pair of vertices $\{u,v\}$.
  In Steps 1(i) to 1(v), the algorithm mentioned in Corollary~\ref{cor:split-all} is applied four times, where 
  each execution takes $O(n^{2R(t, t)+2t+2})$ time (we note that $p=q=t-1$). 
  In Step 1(vi), since there are at most $n^{2R(t,t)}$ $(t-1,t-1)$-split partitions in each list $L_{xy}$ (Lemma~\ref{lem:split-all}),
  Step 1(vi)b is executed at most $n^{8R(t,t)}$ times and each execution of Step 1(vi)b takes $O(f(n)\cdot n^2)$ time.
  Therefore, each iteration of Step 1 has a worst-case time complexity of $O(f(n)\cdot n^{8R(t,t)+2})$.
  Therefore, the algorithm runs in time $O(f(n)\cdot n^{8R(t,t)+4})$ time as there are $O(n^2)$ iterations of Step 1.
  The last statement of the theorem follows from the fact that $K_t$-free graphs can be recognized in time $O(n^t)$.
\end{proof}

Since bounded degenerate graphs have bounded clique number and can be recognized in polynomial-time, 
Theorem~\ref{thm:kt-free} implies that \SCT\ $d$-degenerate
graphs can be solved in polynomial-time, a result obtained in \cite{DBLP:journals/algorithmica/FominGST20}.




\section{Hardness results}
\label{sec:hardness}

In this section, we prove hardness results for \SCT\ $H$-free graphs when $H$ is a star, a path, or a cycle.
Specifically, we prove that, if $H$ is a star with at least six vertices, or a path with at least seven vertices,
or a cycle with at least eight vertices, the problem is NP-complete. For all these hard cases, we prove something
stronger: these hard problems cannot be solved in time $2^{o(|V(G)|)}$, assuming the ETH. This gives us the optimal 
complexity of these problems as these problems can be trivially solved in time $2^{O(|V(G)|)}$.

The proofs are by induction on the number of vertices in $H$.
For base cases, we give reductions from variants of $k$-\textsc{SAT} problem (for example, from \FSAT\ to \SCT\ $K_{1,5}$-free graphs).
For the inductive step, we give reductions from problems of the same kind (for example, from \SCT\ $K_{1,t}$-free graphs to 
\SCT\ $K_{1,t+1}$-free graphs).


\subsection{Stars}
\label{k1t-free}

Construction~\ref{starconstruct} will be used for an \textit{inductive} reduction for \SCT\ $K_{1,t}$-free graphs.

\begin{construction}
\label{starconstruct}
    Let $(G', t)$ be the input to the construction, where $G'$ is a graph and $t\geq 1$ is an integer. 
    For every vertex $u$ of $G'$, 
    introduce $(t+2)$  vertices denoted by the set $W_u$, which includes a special vertex $u'$. 
    Each of these sets induces  a clique ($K_{t+2}$).
    Further, every vertex $u\in V(G')$ is adjacent to  every vertex in $W_u$ except $u'$. 
    Let the resultant graph be $G$ and let $W$, which induces a cluster graph, be the union of all newly introduced vertices.  
\end{construction}

\begin{figure}[ht]
  \centering
    \centering
    \begin{tikzpicture}[myv/.style={circle, draw,color=white, inner sep=2.5pt}, myv1/.style={circle, draw, inner sep=0.5pt},myv2/.style={rectangle, draw,inner sep=16.5pt}, myv3/.style={ellipse, draw},myv4/.style={rectangle, draw, inner sep=1.5pt}, myv5/.style={circle, draw, dashed, inner sep=1.5pt},myv6/.style={circle, draw, inner sep=2.5pt}]

   \node[myv6] (wa) at (170:2.7) {};
   \node[myv6] (wb) at (120:2.5) {};
   \node[myv6] (wc) at (70:2.5) {};
   \node[myv6] (wd) at (10:2.7) {};
   \node[myv6] (wu) at (-90:1.5) {};

  \node[myv4] (wa1) at (170:2.1) {$K_5$};
  \node[myv4] (wb1) at (120:2) {$K_5$};%
  \node[myv4] (wc1) at (70:2) {$K_5$};
  \node[myv4] (wd1) at (10:2.1) {$K_5$};
  \node[myv4] (wu1) at (-90:1) {$K_5$};

  \node[myv3][fit=(wa) (wb) (wc) (wd) (wu) (wa1) (wb1) (wc1) (wd1) (wu1)]{}; 
 
  \node [myv1](u) at (0,0) {u};
  \node[myv1] (a) at (170:0.75) {$x_1$};
  \node[myv1] (b) at (120:0.75) {$x_2$};
  \node[myv1] (c) at (70:0.75) {$x_3$};
  \node[myv1] (d) at (10:0.75) {$x_t$};
  \node [myv] [label=above: $G'$](g) at (-30:2) {};
  \node [myv] [label=above: $G$](g) at (-30:4) {};
  
  \node[myv3][fit=(a) (b) (c) (d)]{}; 
\node[myv5][fit=(wa) (wa1), label=above:$W_{x_1}$] {};
\node[myv5][fit=(wb) (wb1), label=above:$W_{x_2}$] {};

\node[myv5][fit=(wc) (wc1), label=above:$W_{x_3}$] {};

\node[myv5][fit=(wd) (wd1), label=above:$W_{x_4}$] {};
\node[myv5][fit=(wu) (wu1), label=below:$W_{u}$] {};

  \draw [line width=0.5mm](u) -- (a); 
  \draw [line width=0.5mm](u) -- (b);
  \draw [line width=0.5mm](u) -- (c);
  \draw [line width=0.5mm](u) -- (d);
  \draw [line width=0.5mm](wu1) -- (u);
  \draw [line width=0.5mm](wa1) -- (a);
  \draw [line width=0.5mm](wb1) -- (b);
  \draw [line width=0.5mm](wc1) -- (c);
  \draw [line width=0.5mm](wd1) -- (d);
  \draw [line width=0.5mm](wu1) -- (wu);
  \draw [line width=0.5mm](wa1) -- (wa);
  \draw [line width=0.5mm](wb1) -- (wb);
  \draw [line width=0.5mm](wc1) -- (wc);
  \draw [line width=0.5mm](wd1) -- (wd);

\end{tikzpicture}
    \caption{An example of Construction \ref{starconstruct} for $t=4$. The lines connecting a circle and a rectangle indicate that the vertex corresponding to the circle is adjacent to all vertices in the  rectangle.} 
    \label{k1t}
  \end{figure}
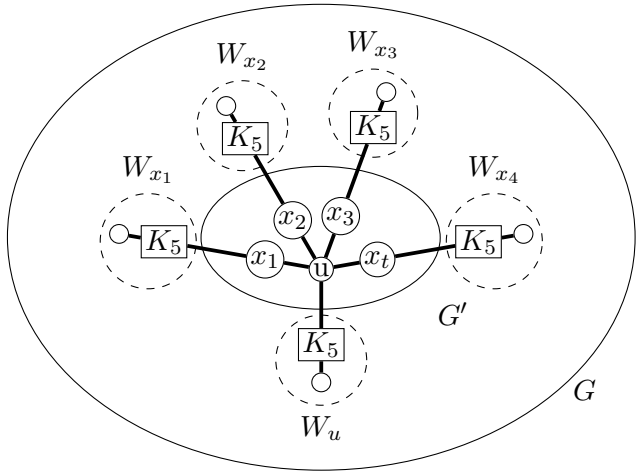

An example of the construction is shown in Figure~\ref{k1t}.

\begin{lemma}
   \label{lem:inductive:star}
   Let $\mathcal{G}'$ and $\mathcal{G}$ be the classes of $K_{1,t}$-free graphs and $K_{1,t+1}$-free graphs respectively for any $t\geq 2$.
   If \SCGD\ is NP-complete, then so is \SCG. Further, if \SCGD\ cannot be solved in time $2^{o(|V(G)|)}$, then so is \SCG.
\end{lemma}
\begin{proof}
  Let $G'$ be an instance of \SCGD. Apply Construction~\ref{starconstruct} on $(G', t)$ to obtain a graph $G$. 
  Clearly, $G$ has $|V(G')|+|V(G')|\cdot (t+2)$ vertices, and hence the reduction is a linear reduction. 
  Now, it is sufficient to prove that $G'$ is a yes-instance of \SCGD\ if and only if $G$ is a yes-instance of \SCG.
  
  Let $G'$ be a yes-instance of \SCGD. Let $S'$ be a solution, i.e., $G'\oplus S'$ is $K_{1,t}$-free. 
  We claim that $G\oplus S'$ is $K_{1,t+1}$-free.
  Assume for a contradiction that there is a $K_{1,t+1}$ induced by a set $F$ of vertices in $G\oplus S'$. 
  Let $u$ be the center of the star $K_{1,t+1}$ induced by $F$ in $G\oplus S'$.
  Since the neighborhood of every vertex in $W$ induces a graph with independence number at most 2, $u\notin W$.
  Therefore, $u\in V(G)\setminus W$.
  Since $u$ is adjacent to only $W_u\setminus \{u'\}$ in $W$, and $W_u\setminus \{u'\}$ forms a clique, we obtain that 
  $|F\cap W|\leq 1$. Therefore, $G'\oplus S'$ has a $K_{1,t}$ as an induced subgraph, which is a contradiction.
  
    For the other direction, assume that $G$ is a yes-instance of \SCG. 
    Then, there is a set $S\subseteq V(G)$ such that $G\oplus S$ is $K_{1,t+1}$-free.
  We claim that $G'\oplus S$ is $K_{1,t}$-free. 
  For a contradiction, assume that there is a set $F$ of vertices which induces $K_{1,t}$ 
  in $G'\oplus S$.  
  Let $u$ be the center of the star induced by $F$.
  If there is a vertex $y$ in $W_u\setminus \{u'\}$ which is not in $S$, then $F\cup \{y\}$ induces a $K_{1,t+1}$ in $G\oplus S$, which is a contradiction.
  Therefore, $W_u\setminus \{u'\}\subseteq S$. 
  Then, if $u\notin S$, then $(W_u\setminus \{u'\})\cup \{u\}$ induces a $K_{1,t+1}$ in $G\oplus S$, which is a contradiction.
  Therefore, $u\in S$.
  Similarly, if $u'\notin S$, then $W_u$ induces a $K_{1,t+1}$ in $G\oplus S$, which is a contradiction.
  Therefore, $u'\in S$.
  Further, if there is a vertex $y\in F\setminus \{u\}$ which is in $S$, then $(W_u\setminus \{u'\})\cup \{y\}$ induces a $K_{1,t+1}$ in $G\oplus S$,
  which is a contradiction.
  Therefore, only $u$ in $F$ is in $S$.
  Then we obtain that $F\cup \{u'\}$ induces a $K_{1,t+1}$ in $G\oplus S$, which is a contradiction.
\end{proof}

\begin{figure}[ht]
  \centering
    \centering
    \begin{tikzpicture}  [myv/.style={circle, draw, inner sep=0pt},myv1/.style={rectangle, draw,inner sep=1pt},myv2/.style={rectangle, draw,inner sep=2.5pt},my/.style={rectangle, draw,dotted,inner sep=0pt},myv3/.style={circle, draw, inner sep=1.5pt}] 
 
\node[myv2]  (a1) at (0.5,4.5) {$V_1$};
\node[myv2]  (a2) at (5.25,4.5) {$V_2$};
\node[myv2]  (a3) at (10.5,4.5) {$V_3$};

   \node[myv3] (x1) at (-0.5,2.65) {$u_{1}$};
   \node[myv] (x1') at (0.9,2.65) {$u'_{1}$};
   \node[myv3] (x2) at (2.2,2.65) {$u_{2}$};
   \node[myv] (x2') at (3.6,2.65) {$u'_{2}$};
   \node[myv3] (x3) at (4.95,2.65) {$u_{3}$};
   \node[myv] (x3') at (6.3,2.65) {$u'_{3}$};
   \node[myv3] (x4) at (7.6,2.65) {$u_{4}$};
   \node[myv] (x4') at (8.9,2.65) {$u'_{4}$};
   \node[myv3] (x5) at (10.2,2.65) {$u_{5}$};
   \node[myv] (x5') at (11.5,2.65) {$u'_{5}$};

    \node[myv1]  (y1) at (0.2,2) {$U_{1,1}$};
    \node[myv1]  (y2) at (-0.5,1.25) {$U_{1,2}$};
    \node[myv1]  (y3) at (0.9,1.25) {$U_{1,4}$};
    \node[myv1]  (y4) at (0.2,0.65) 
    {$U_{1,3}$};

    \node[myv1]  (y5) at (2.9,2) {$U_{2,1}$};
    \node[myv1]  (y6) at (2.2,1.25) {$U_{2,2}$};
    \node[myv1]  (y7) at (3.6,1.25) {$U_{2,4}$};
    \node[myv1]  (y8) at (2.9,0.65) {$U_{2,3}$};

    \node[myv1]  (y9) at (5.65,2) {$U_{3,1}$};
    \node[myv1]  (y10) at (4.95,1.25) {$U_{3,2}$};
     \node[myv1]  (y11) at (6.3,1.25) {$U_{3,4}$};
    \node[myv1]  (y12) at (5.65,0.65) {$U_{3,3}$};

    \node[myv1]  (y13) at (8.2,2) {$U_{4,1}$};
    \node[myv1]  (y14) at (7.6,1.25) {$U_{4,2}$};
    \node[myv1]  (y15) at (9,1.25) {$U_{4,4}$};
    \node[myv1]  (y16) at (8.2,0.65) {$U_{4,3}$};
      
    \node[myv1]  (y17) at (10.9,2) {$U_{4,1}$};
    \node[myv1]  (y18) at (10.4,1.25) {$U_{4,2}$};
    \node[myv1]  (y19) at (11.8,1.25) {$U_{4,4}$};
     \node[myv1]  (y20) at (10.9,0.65) {$U_{4,3}$};

  
 \node[my][fit=(a1) (a3),  inner xsep=1.5ex, inner ysep=3.5ex, label=above:$V'$] {};




  \draw (a1) -- (x1);
  \draw (a1) -- (x2);
  \draw (a1) -- (x3);
  \draw (a1) -- (x4);
  
  \draw (a2) -- (x1');
  \draw (a2) -- (x2');
  \draw (a2) -- (x3');
  \draw (a2) -- (x5);
  
  \draw (a3) to[bend right=7] (x1);
  \draw (a3) -- (x2);
  \draw (a3) -- (x4');
  \draw (a3) -- (x5);
  
  \draw (a1) -- (a2);
  \draw (a1) to[bend left=10] (a3);
  \draw (a2) -- (a3);

 \draw (x1) -- (x1');
 \draw (x1) -- (y1);
 \draw (x1') -- (y1);
  \draw (y1) -- (y2);
  \draw (y1) -- (y3);
  \draw (y1) -- (y4);
  
   \draw (x2) -- (x2');
   \draw (x2) -- (y5);
   \draw (x2') -- (y5);
   \draw (y5) -- (y6);
   \draw (y5) -- (y7);
   \draw (y5) -- (y8);
   
   \draw (x3) -- (x3');
   \draw (x3) -- (y9);
   \draw (x3') -- (y9);
   \draw (y9) -- (y10);
   \draw (y9) -- (y11);
   \draw (y9) -- (y12);
   
    \draw (x4) -- (x4');
   \draw (x4) -- (y13);
   \draw (x4') -- (y13);
   \draw (y13) -- (y14);
   \draw (y13) -- (y15);
   \draw (y13) -- (y16);
   
   \draw (x5) -- (x5');
   \draw (x5) -- (y17);
   \draw (x5') -- (y17);
   \draw (y17) -- (y18);
   \draw (y17) -- (y19);
   \draw (y17) -- (y20);


\end{tikzpicture}  
    \caption{An example of Construction \ref{constructionk15} for the formula 
    $\phi = C_1 \wedge C_2\wedge C_3$ where $C_1$= $x_1 \lor x_2 \lor {x_3} \lor {x_4}$, 
    $C_2= \overline{x_1} \lor \overline{x_2} \lor \overline{x_3} \lor {x_5}$, and 
    $C_3 = x_1 \lor x_2 \lor \overline{x_4} \lor x_5 $.  
    Each rectangle represents a $K_5$. 
    The lines connecting two rectangles indicate that each vertex in one rectangle is adjacent to all vertices in the other rectangle. 
    Lines connecting a rectangle and a vertex have a similar meaning.}
    \label{k15fig}
  \end{figure}
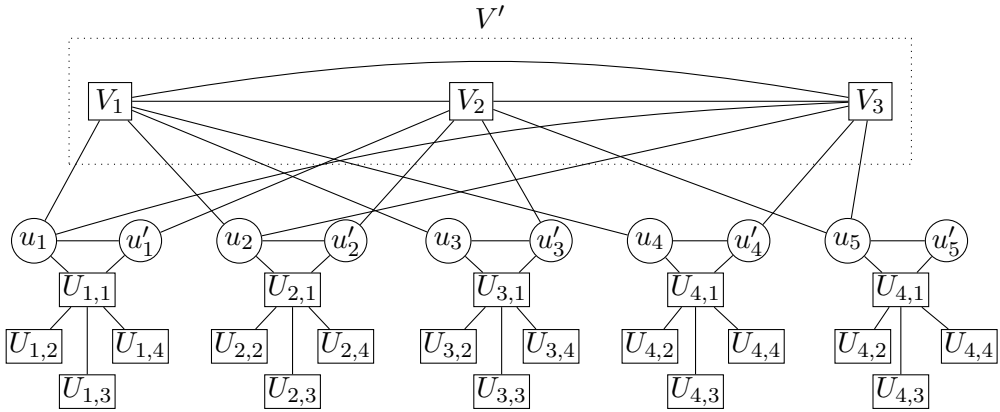

Now, to utilize Lemma~\ref{lem:inductive:star}, we need a hardness result when $H = K_{1,t}$ for some $t\geq 2$, 
the smaller the $t$, the better the implications will be. We obtain such a result for $t=5$ using Construction~\ref{constructionk15}.
\begin{construction}
\label{constructionk15}
    Let $\Phi$ be the input to the construction, where $\Phi$, with $n$ variables and $m$ clauses,  is a \FSATO\ formula.
    We construct $G=(V,E)$ 
    in the following way:
    
\begin{itemize}
    \item For each variable $X_i$ in $\Phi$, introduce two vertices - one vertex, denoted by $u_i$, 
    for the positive literal $x_i$  and one vertex, denoted by $u'_i$, 
    for the negative literal $\overline{x_i}$. The vertex $u_i$ is adjacent to $u'_i$. 
    Further, for each variable $X_i$ introduce four sets $U_{i,1}$, $U_{i,2}$, $U_{i,3}$ and $U_{i,4}$ of five vertices each.  
    Each of these sets  induces a $K_5$. 
    The vertices $u_i$ and $u'_i$ are all-adjacent to $U_{i,1}$. 
    Further, $U_{i,1}$ is all-adjacent to $U_{i,2}$, $U_{i,3}$ and $U_{i,4}$. 
    Thus, the total number of vertices corresponding to a variable of $\Phi$ is 22.
    
    \item  For each clause $C_i$ of the form  $\ell_{i,1}\lor \ell_{i,2}\lor \ell_{i,3}\lor \ell_{i,4}$ 
    introduce a set $V_{i}$ of five vertices each which 
    induces a $K_5$. All the vertices of each $V_i$  together form a big clique, denoted by $V'$, of size 5$\cdot m$. 
    Let the four vertices introduced (in the previous step) for the literals $\ell_{i,1}, \ell_{i,2}, \ell_{i,3}$,  and $\ell_{i,4}$ be denoted by $y_{i,1},y_{i,2},y_{i,3}$ and $y_{i,4}$ respectively. 
    We note that, if $\ell_{i,1} = x_j$, then $y_{i,1} = u_j$, and if $\ell_{i,1} = \overline{x_j}$, then $y_{i,1} = u'_j$.
    Similarly, let the four vertices introduced for the negation of these literals be denoted by  
    $z_{i,1},z_{i,2},z_{i,3}$, and $z_{i,4}$  respectively. 
    We note that, if $\ell_{i,1} = x_j$, then $z_{i,1} = u'_j$, 
    and if $\ell_{i,1} = \overline{x_j}$, then $z_{i,1} = u_j$.
    Further, every vertex in $V_{i}$ is adjacent to the vertices $y_{i,1}$, $y_{i,2}$, $y_{i,3}$ and $y_{i,4}$. 
    
    
\end{itemize}
This completes the construction (refer Figure \ref{k15fig}).
\end{construction}

For convenience, we call each set $V_i$  introduced in the construction a \textit{clause set} with each of them containing five vertices. 
For each variable $X_i$, the vertices $u_i$ and $u'_i$ are called \textit{literal vertices}.
The union of all literal vertices is denoted by $U$.
Further, for each variable $X_i$, the sets $U_{i,s}$ 
(for $1\leq r\leq 4$) are called \textit{hanging sets} 
and the union of which is denoted by $U^h_i$. By $U^h$, we denote the union of all $U^h_i$s.
The vertices in the hanging sets are called \textit{hanging vertices}.

Construction~\ref{constructionk15} will be used for a reduction from \FSAT\ to \SCT\ $K_{1,5}$-free graphs.
Whenever we introduce a $K_5$ in the construction, the objective is to forbid a solution $S$ to have all
the vertices of that $K_5$. If all the five vertices in a $K_5$ is in $S$, then those vertices along with 
a vertex adjacent to them (but not in $S$) or a vertex nonadjacent to them (but in $S$) form a $K_{1,5}$
in $G\oplus S$, where $G$ is the output of the construction. This makes sure that both $u_i$ and $u'_i$
are not in $S$ together (otherwise, they form a $K_{1,5}$ along with their hanging vertices not in $S$ -- note that
each hanging set forms a $K_5$). 
Further, for every $C_i$, at least two literal vertices corresponding to the literals
in $C_i$ must be in $S$, otherwise there will be $K_{1,5}$ where the center vertex is a vertex from $V_i$.
These observations help us to get a valid truth assignment which satisfies the formula $\Phi$, an instance of \FSAT.

\begin{theorem}
\label{thm:k15}
    \SCT\ $K_{1,5}$-free graphs is NP-complete. Further, the problem cannot be solved in time $2^{o(|V(G)|)}$, assuming the ETH.
\end{theorem}

\begin{proof}
     Let $G$ be a graph obtained by Construction \ref{constructionk15} on the input $\Phi$, 
     where $\Phi$, with $n$ variables and $m$ clauses, is an instance of \FSAT\ problem. 
     It can be observed that $G$ has $22n+5m$ vertices. Therefore, the reduction is a linear reduction.
     Now, It is sufficient to prove that $\Phi$ is a \textit{yes}-instance of \FSAT\ problem if and only if 
     there exists a set $S\subseteq V(G)$ of vertices such that $G\oplus S$ is $K_{1,5}$-free. 

    To prove the forward direction, assume that $\Phi$ is satisfiable and let $L$ be the set of true literals. 
    We note that for every variable $X_i$, either $x_i$ or $\overline{x_i}$ is true but not both. 
    Let $S$ be the set of all literal vertices corresponding to the true literals, i.e,. if $x_i$ is true then $u_i$ is in $S$,
    otherwise, the vertex $u'_i$ is in $S$. 
    Our claim is that $G\oplus S$ is $K_{1,5}$-free. 
    For a contradiction, assume that $G\oplus S$ contains $K_{1,5}$ induced by $W\subseteq V(G)$.
    
    Claim 1: For every $i$ (for $1\leq i\leq m$), $|V_{i}\cap W| \leq 1$
    
    Proof of Claim 1: 
    For a contradiction, assume that there are two vertices $a,b \in V_{i}\cap W$. 
    Since $V_i$ induces a clique, $a$ and $b$ are adjacent.
    Then one of them must be the center vertex and the other must be one of the leaf vertices of the $K_{1,5}$ induced by $W$.
    Then we get a contradiction as $V_i$ is a module and the neighborhood of a center vertex and a leaf vertex cannot be the same in $K_{1,5}$.
    
    Claim 2:  For every $i,s$ (for which $U_{i,s}$ is defined), $|U_{i,s}\cap W| \leq 1$. 
    
    We omit the proof of Claim 2 as it is similar to that of Claim 1.
    
    Claim 3: The set $U\cap W$ induces an empty graph (in $G\oplus S$) with at most five vertices, or  a $K_2$, or a $K_{1,2}$. 
    
    Proof of Claim 3: 
    We analyse the cases based on the number of edges in the graph induced by $U\cap W$.
    Case (i): The set $U\cap W$ induces a graph with no edges 
    -- then it is an empty graph with at most five vertices as there is no independent set of size at least six in a $K_{1,5}$.
    Case (ii): The set $U\cap W$ induces a graph with exactly one edge. 
    Then, clearly, it is of the form $K_2$ as there is no induced $K_2\cup K_1$ in a $K_{1,5}$.
    Case (iii): The set $U\cap W$ induces a graph with exactly two edges. 
    Clearly, it is of the form $K_{1,2}$ as there is no induced $2K_2$ or $K_{1,2}\cup K_1$ in $K_{1,5}$. 
    Case (iv): The set $U\cap W$ induces a graph with at least three edges.
    Then the graph induced by $U\cap W$ must be $K_{1,s}$ for some $s\geq 3$.
    In $G\oplus S$, all the vertices corresponding to true literals form a clique and vertices corresponding to 
    true literals are adjacent to vertices corresponding to their false literals. 
    Therefore, the graph induced by $U$ in $G\oplus S$ does not have an induced claw.
    This leads to a contradiction. 

     With these claims, we are ready to complete the proof of the forward direction. 
     By Claim 3, the vertices in $U$ cannot induce a $K_{1,5}$, therefore either $V'\cap W$ or $U^h\cap W$ is nonempty. 
     
      Case 1: $V'\cap W=\emptyset$. 
     Since the neighborhood of any hanging vertex induces a graph with independence number at most four,
     the center vertex of the $K_{1,5}$ must be from $U$. Without loss of generality, assume that the center 
     vertex is $u_i$ for some variable $X_i$. Claim 3 implies that at least three leaf vertices of the 
     $K_{1,5}$ must be from $U^h$.
     By Claim 2, $W$ has only at most one vertex from $U_{i,1}$. 
     Since $u_i$ is nonadjacent to other hanging sets, we obtain a contradiction.

     Case 2: $|V'\cap W| = 1$.  Let the vertex $v_i$ in $V'\cap W$ be from a clause set $V_i$.
     There can be two cases: (i) $|U^h\cap W| =0$; (ii) $|U^h\cap W|\geq 1$. 
     First we prove the case in which $|U^h\cap W| =0$ and then we prove $|U^h\cap W| \geq 1$. 
     Assume that $|U^h\cap W| =0$. Then $U\cap W$
     induces  $I_5$ 
     or $K_{1,4}$. As per Claim 3, the latter gives a contradiction. 
     Therefore, $U\cap W$ induces $I_5$ and $v_i$ is the center vertex of the $K_{1,5}$, 
     i.e., $v_i$ is adjacent to all the five vertices in $U\cap W$.
     This gives a contradiction as $v_i$ is adjacent to only four vertices in $U$.
     Now, assume that $|U^h\cap W|\geq 1$. 
     We note that the sets $U^h$ and $V'$ are nonadjacent. 
     Therefore, $\{v_i\}\cup (W\cap U^h)$ are leaf vertices of the $K_{1,5}$ induced by $W$ in $G\oplus S$.
     Therefore, the center vertex of the $K_{1,5}$ must be a literal vertex $y_{i,s}$ (for some $1\leq s\leq 4$).
     Let $y_{i,s}$ corresponds to a variable $X_j$. 
     Then, $U^h\cap W$ contains only a single vertex from $U_{j,1}$ (See Claim 2) and no vertices from other hanging sets ($y_{i,s}$
     is not adjacent to them).
     Then three leaf vertices of the $K_{1,5}$ must be from $U$. Then $y_{i,s}$
     along with these three vertices induces a $K_{1,3}$ which is a contradiction as per Claim 3.
    
     Case 3: $|V'\cap W| \geq 2$. 
     Since $V'$ induces a clique, $|V'\cap W| = 2$ and $V'\cap W$ induces a $K_2$. 
     Let the vertices in $K_2$  be $v_i\in V_i$ and $v_j\in V_j$.
     By Claim 1, $i\neq j$. 
     Clearly, one of the vertex in $\{v_i,v_j\}$, say $v_i$, must be the center vertex of the $K_{1,5}$.
     Therefore, all other four vertices, which forms a $I_4$, of the $K_{1,5}$ must be from $U$.
     That implies, all the four literal vertices $\{y_{i,1}, y_{i,2}, y_{i,3}, y_{i,4}\}$ corresponding to the literals in $C_i$ are in $W$.
     Since at least two literals are true in $C_i$, at least two of these literal vertices are in $S$ 
     and these four vertices does not induced a $I_4$, which is a contradiction.

     
     To prove the other direction, assume that $G$ is a yes-instance, and $S\subseteq V(G)$ is a solution, i.e., 
     $G\oplus S$ is $K_{1,5}$-free. We claim that $\Phi$ is satisfiable with the truth assignment for $\Phi$ as follows:

\begin{center}
 \begin{math} X_i=\left\{\begin{array}{ll}True, & \mbox{if the vertex $u_i$ is in $S$}.\\
      False, & \mbox{otherwise}.
    \end{array}
  \right.
\end{math}
\end{center}

    For a contradiction, assume that $\Phi$ is not satisfiable.
    
    Claim 4: The set $S$ 
    does not contain all vertices of $V_{i}$ for any $i$ (for $1\leq i\leq m$).
    
    Proof of Claim 4: Assume that all the vertices in $V_{i}$ (for any $i$) are in $S$. 
    Assume that there exists a vertex $v\in V'\setminus V_i$ such that $v\notin S$.
    Then $V_i\cup \{v\}$ induces a $K_{1,5}$ in $G\oplus S$, which is a contradiction.
    Therefore, $V'\subseteq S$. 
    Assume that there is a literal vertex, say $y_{i,s}$, which is not in $S$. Then $V_i\cup \{y_{i,s}\}$
    induces a $K_{1,5}$ in $G\oplus S$, which is a contradiction.
    This implies that every literal vertex, such that the corresponding literal appears in any of the clauses,
    is in $S$.
    Let $X_j$ be a variable which does not appear in $C_i$ 
    (we can safely assume that such a variable exists, otherwise the formula has only four variables).
    Clearly, either $u_j$ or $u'_j$ appears in some clauses. Therefore, one of them is in $S$.
    Without loss of generality, assume that $u_j$ is in $S$.
    Then $V_i\cup \{u_j\}$ induces a $K_{1,5}$ in $G\oplus S$, which is a contradiction.
    
    Claim 5: Let $U_{i,s}$ (for any $1\leq s\leq 4$) be a hanging set corresponding to a variable $X_i$.
    Then there is at least one vertex in $U_{i,s}$ not in $S$.
    
    Proof of Claim 5: Assume for the contradiction that $U_{i,s}\subseteq S$.
    We need to analyse the case when $s=1$ and $s=2$, the cases $s=3$ and $s=4$ are 
    analogous to the case $s=2$.
    Assume that $s=1$. Since all vertices in $U_{i,1}$ is in $S$, all vertices in 
    $U_{i,2}, U_{i,3}$ and $U_{i,4}$ must be in $S$, otherwise there will be an induced $K_{1,5}$ formed
    by vertices in $U_{i,1}$ and a vertex in $U_{i,2}\cup U_{i,3}\cup U_{i,4}\setminus S$.
    Now, $U_{i,2}$ along with any vertex in $U_{i,3}$ induces a $K_{1,5}$ in $G\oplus S$, which is a contradiction.
    Now, we verify the claim for $s=2$.
    For a contradiction, assume that $U_{i,2}\subseteq S$. 
    Then all vertices in $U_{i,1}$ is in $S$, which is a contradiction by the previous case.
    
    Claim 6: If a literal vertex is in $S$, then the corresponding literal is assigned true by $\Phi$.
    
    Proof of Claim 6:
    We observe that both literal vertices, $u_i$ and $u'_i$, of a variable $X_i$
    cannot be in $S$ together. If that happens, then $u_i$, $u'_i$, and exactly one vertex each from
    $U_{i,1}\setminus S$, $U_{i,2}\setminus S$, $U_{i,3}\setminus S$, and $U_{i,4}\setminus S$ (by Claim 5, these sets are nonempty)
    induce a $K_{1,5}$. This implies that if a literal vertex is in $S$, the corresponding literal is assigned true. 
    
    With these claims, we are ready to complete the proof of the backward direction. 
    We need to prove that, for every clause $C_i$, at least two literals are true.
    For a contradiction, assume that there exists a clause $C_i$ in which only at most one literal is true.
    Then by Claim 6, only at most one vertex in $\{y_{i,1}, y_{i,2}, y_{i,3}, y_{i,4}\}$ is in $S$.
    We can safely assume that there is a clause $C_j$ which does not contain any of the literals in $C_i$ (if this is not the case,
    then we can add a dummy clause with literals of new variables and do the reduction from the new formula).
    By Claim 4, at least one vertex, say $v_i\in V_i$ and at least one vertex, say $v_j\in V_j$ are not in $S$.
    Now, $\{v_i, v_j, y_{i,1}, y_{i,2}, y_{i,3}, y_{1,4}\}$ induces a $K_{1,5}$ in $G\oplus S$, which is a contradiction.
%
\end{proof}

Theorem~\ref{thm:k1t} is a direct implication of Lemma~\ref{lem:inductive:star} and Theorem~\ref{thm:k15}.
\begin{theorem}
\label{thm:k1t}
Let $t\geq 5$ be any integer. Then \SCT\ $K_{1,t}$-free graphs  is NP-complete.
Further, the problem cannot be solved in time $2^{o(|V(G)|)}$, unless the ETH fails. 
\end{theorem}



\subsection{Paths}
\label{pt-free}
Here, we obtain hardness results for \SCT\ $H$-free graphs when $H$ is a path.
We start with a construction which will be used for an inductive reduction from \SCT\ $P_t$-free graphs to \SCT\ $P_{t+2}$-free graphs.
\begin{construction}
\label{pathconstruct}
    Let $(G', t)$ be the input to the construction, where $G'$ is a graph and $t\geq 1$ is an integer. 
    For every vertex $u$ of $G'$, 
    introduce $(t+2)$  vertices denoted by the set $W_u$. 
    Each of these sets induces  a  $\overline{P_{t+2}}$.
    Further, every vertex $u\in V(G')$ is adjacent to  every vertex in $W_u$. 
    Let the resultant graph be $G$ and let $W$, which induces a disjoint union of $\overline{P_{t+2}}$, be the union of all newly introduced vertices.  
\end{construction}

\begin{figure}[ht]
  \centering
    \centering
    \begin{tikzpicture}[myv/.style={circle, draw,color=white, inner sep=2.5pt}, myv1/.style={circle, draw, inner sep=2.5pt},myv2/.style={rectangle, draw,inner sep=16.5pt}, myv3/.style={ellipse, draw, inner sep=8.5pt},myv4/.style={rectangle, draw, inner sep=2.5pt}, myv5/.style={circle, draw, inner sep=4pt}]

  \node[myv4] [label=above: $W_{x_1}$](wa1) at (140:2.2) {$\overline{P_{7}}$};
  \node[myv4] [label=above: $W_{u}$] (wb1) at (160:3.4) {$\overline{P_{7}}$};%
  \node[myv4]  [label=above: $W_{x_3}$] (wc1) at (35:2.4) {$\overline{P_{7}}$};
  \node[myv4]  [label=above: $W_{v}$](wd1) at (20:3.4) {$\overline{P_{7}}$};
  \node[myv4] [label=above: $W_{x_2}$] (wu1) at (90:1.5) {$\overline{P_{7}}$};
  \node[myv] (v) at (-90:0.5) {};

 
 \node[myv3] [label=right: $G$] [fit= (a) (b) (c) (d) (v) (wa1) (wb1) (wc1) (wd1) (wu1)]{}; 
 
  \node[myv5] (b) at (180:2) {u};
  \node[myv1] (a) at (180:1) {$x_1$};
   \node [myv1](u) at (0,0) {$x_2$};
  \node[myv1] (c) at (180:-1) {$x_3$};
  \node[myv5] (d) at (180:-2) {v};
  \node [myv] (v) at (-90:0.25) {};
  
  \node[myv3] [label=right: $G'$] [fit=(a) (b) (c) (d) (v)]{};

  \draw [line width=0.5mm](u) -- (a); 
  \draw [line width=0.5mm](a) -- (b);
  \draw [line width=0.5mm](u) -- (c);
  \draw [line width=0.5mm](c) -- (d);
  \draw [line width=0.5mm](wu1) -- (u);
  \draw [line width=0.5mm](wa1) -- (a);
  \draw [line width=0.5mm](wb1) -- (b);
  \draw [line width=0.5mm](wc1) -- (c);
  \draw [line width=0.5mm](wd1) -- (d);

\end{tikzpicture}
    \caption{An example of Construction \ref{pathconstruct} for $t=5$. The lines connecting a circle and a rectangle indicate that the vertex corresponding to the circle is adjacent to all vertices in the  rectangle.} 
    \label{pt}
  \end{figure}
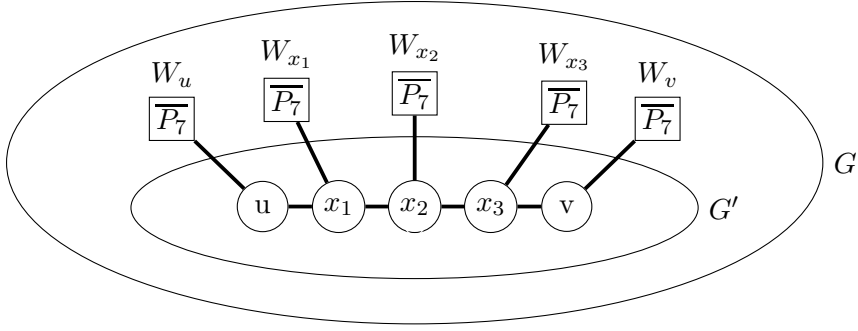

An example of the construction is shown in Figure~\ref{pt}.

\begin{lemma}
   \label{lem:inductive:path}
   Let $\mathcal{G}'$ and $\mathcal{G}$ be the classes of $P_t$-free graphs and $P_{t+2}$-free graphs respectively for any $t\geq 3$.
   If \SCGD\ is NP-complete, then so is \SCG. Further, if \SCGD\ cannot be solved in time $2^{o(|V(G)|)}$, then so is \SCG.
\end{lemma}
\begin{proof}
  Let $G'$ be an instance of \SCGD. Apply Construction~\ref{pathconstruct} on $(G', t)$ to obtain a graph $G$. 
  Clearly, $G$ has $|V(G')|+|V(G')|\cdot (t+2)$ vertices, and hence the reduction is a linear reduction. 
  Now, it is sufficient to prove that $G'$ is a yes-instance of \SCGD\ if and only if $G$ is a yes-instance of \SCG.
  
  Let $G'$ be a yes-instance of \SCGD. Let $S'$ be a solution, i.e., $G'\oplus S'$ is $P_t$-free. 
  We claim that $G\oplus S'$ is $P_{t+2}$-free.
  Assume for a contradiction that there is a $P_{t+2}$ induced by a set $F$ of vertices in $G\oplus S'$. 
  
  Claim 1: For every vertex $u\in V(G')$, $|W_u\cap F|\leq 1$.
  
  Proof of Claim 1:
  For a contradiction, assume that $|W_u\cap F|\geq 2$.
  Since $P_{t+2}$ is not isomorphic to $\overline{P_{t+2}}$ for $t\geq 3$, we obtain that at least one vertex in
  $F$ is from outside $W_u$. Since $P_{t+2}$ is a connected graph and $W_u$ is adjacent only to $u$,
  we obtain that $u\in F$. Then, only at most two vertices in $W_u$ are in $F$, otherwise there will be a 
  $K_{1,3}$ as a subgraph (not necessarily induced) formed by $u$ and any three vertices in $F\cap W_u$. This
  is a contradiction, as there is no $K_{1,3}$ as a subgraph in $P_{t+2}$.
  Assume that there are two vertices $a,b$ in $F\cap W_u$. Then $a$ and $b$ must be nonadjacent.
  Otherwise, $u,a,b$ forms a triangle. Therefore, $a$ and $b$ are the end vertices 
  of the $P_{t+2}$ induced by $F$. Since $u$ is a common neighbor of $a$ and $b$, we obtain 
  a contradiction, as there is no common neighbor for end vertices of $P_{t+2}$
  for $t\geq 3$.
  
  Claim 1 implies that if $F\cap W_u$ is nonempty, then it contains exactly one vertex which is an end vertex
  of the $P_{t+2}$. Therefore, all intermediate vertices of the $P_{t+2}$ are from the copy of $G'$ in $G$.
  Therefore, $G'\oplus S'$ contains an induced $P_t$, which is a contradiction.
  
  

    For the other direction, assume that $G$ is a yes-instance of \SCG. Then, there is a set $S\subseteq V(G)$ such that $G\oplus S$ is $P_{t+2}$-free.
    We claim that $G'\oplus S$ is $P_t$-free. 
    For a contradiction, assume that there is a set $F$ of vertices which induces a $P_t$ in $G'\oplus S$. 
    Let $u$ and $v$ be the end vertices  of the path induced by $F$.
   If $S$ contains all vertices in $W_{u}$, then it forms a $P_{t+2}$ in $G\oplus S$, which is a contradiction. 
   Therefore, there is at least one vertex $u'\in W_{u}$, which is not  in $S$. 
   Similarly, there is at least one vertex $v'\in W_{v}$  which is not  in $S$.  
   This gives a contradiction as $F\cup\{u',v'\}$ induces  $P_{t+2}$ in $G\oplus S$.
  \end{proof}


Lemma~\ref{lem:inductive:path} gives us a challenge - in order to obtain hardness results for every path $P_t$, we need to come up with two hardness reductions one for $H=P_t$
and another for $H=P_{t+1}$. We overcome this by proving hardness for $H=P_7$ and $H=P_8$. Construction~\ref{constructionp7} is used for \SCT\ $P_7$-free graphs, a variant of it 
will be used later for \SCT\ $P_8$-free graphs.

\begin{figure}[ht]
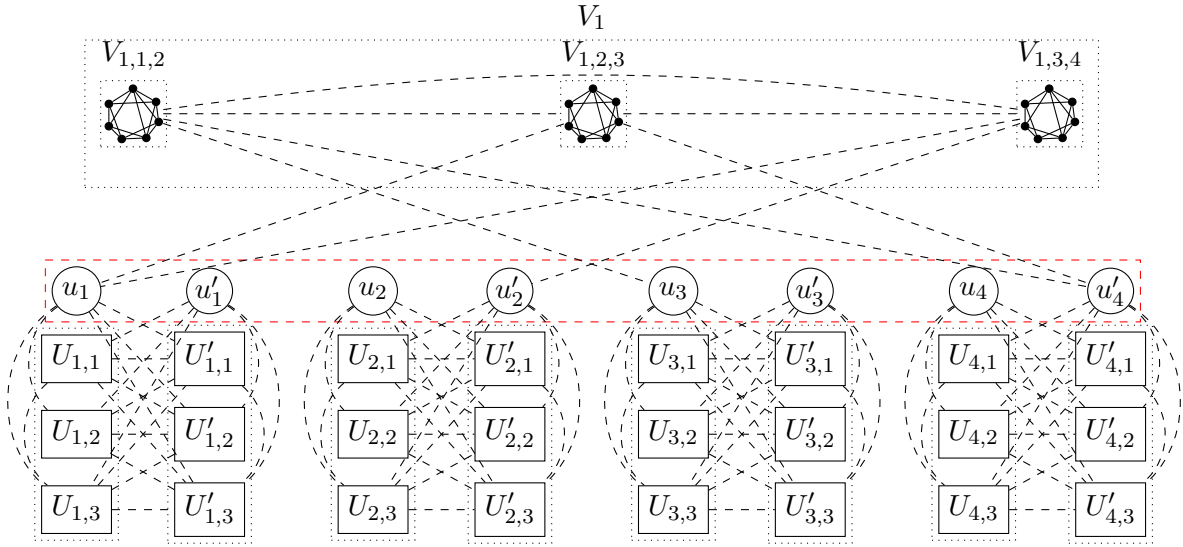

  \centering
    \centering
    \begin{tikzpicture}  [myv/.style={circle, draw, inner sep=2pt},myv1/.style={rectangle, draw},myv2/.style={rectangle, draw,dotted},myv3/.style={circle, draw, inner sep=0.5pt},my/.style={\input{figs/hard/c},draw}, myv4/.style={rectangle, draw,dashed,color=red}] 
 
\node[inner sep=0pt] (a1) at (-0.05,5)
    {\input{figs/hard/p7c}};
\node[inner sep=0pt] (a2) at (6,5) {\input{figs/hard/p7c}};
\node[inner sep=0pt] (a3) at (12,5)
    {\input{figs/hard/p7c}};

   \node[myv] (x1) at (-0.8,2.65) {$u_{1}$};
   \node[myv3] (x1') at (0.95,2.65) {$u'_{1}$};
   \node[myv] (x2) at (3.10,2.65) {$u_{2}$};
   \node[myv3] (x2') at (4.90,2.65) {$u'_{2}$};
   \node[myv] (x3) at (7.05,2.65) {$u_{3}$};
   \node[myv3] (x3') at (8.85,2.65) {$u'_{3}$};
   \node[myv] (x4) at (11,2.65) {$u_{4}$};
   \node[myv3] (x4') at (12.8,2.65) {$u'_{4}$};

    \node[myv1]  (y1) at (-0.8,1.75) {$U_{1,1}$};
    \node[myv1]  (y2) at (-0.8,0.75) {$U_{1,2}$};
    \node[myv1]  (y3) at (-0.8,-0.25) {$U_{1,3}$};
    \node[myv1]  (y4) at (0.95,1.75) {$U'_{1,1}$};
    \node[myv1]  (y5) at (0.95,0.75) {$U'_{1,2}$};
    \node[myv1]  (y6) at (0.95,-0.25) {$U'_{1,3}$};
    
    \node[myv1]  (y7) at (3.10,1.75) {$U_{2,1}$};
    \node[myv1]  (y8) at (3.10,0.75) {$U_{2,2}$};
    \node[myv1]  (y9) at (3.10,-0.25) {$U_{2,3}$};
    \node[myv1]  (y10) at (4.90,1.75) {$U'_{2,1}$};
    \node[myv1]  (y11) at (4.90,0.75) {$U'_{2,2}$};
    \node[myv1]  (y12) at (4.90,-0.25) {$U'_{2,3}$};
    
    \node[myv1]  (y13) at (7.05,1.75) {$U_{3,1}$};
    \node[myv1]  (y14) at (7.05,0.75) {$U_{3,2}$};
    \node[myv1]  (y15) at (7.05,-0.25) {$U_{3,3}$};
     \node[myv1]  (y16) at (8.85,1.75) {$U'_{3,1}$};
    \node[myv1]  (y17) at (8.85,0.75) {$U'_{3,2}$};
    \node[myv1]  (y18) at (8.85,-0.25) {$U'_{3,3}$};
    
     \node[myv1]  (y19) at (11,1.75) {$U_{4,1}$};
    \node[myv1]  (y20) at (11,0.75) {$U_{4,2}$};
    \node[myv1]  (y21) at (11,-0.25) {$U_{4,3}$};
     \node[myv1]  (y22) at (12.8,1.75) {$U'_{4,1}$};
    \node[myv1]  (y23) at (12.8,0.75) {$U'_{4,2}$};
    \node[myv1]  (y24) at (12.8,-0.25) {$U'_{4,3}$};

  
 \node[myv2][fit=(a1) (a3),  inner xsep=1.5ex, inner ysep=3.5ex, label=above:$V_1$] {}; 
 
 
 
 \node[myv2][fit=(a1),  inner xsep=0.25ex, inner ysep=0.25ex, label=above:$V_{1,1,2}$] {}; 
 \node[myv2][fit=(a2),  inner xsep=0.25ex, inner ysep=0.25ex, label=above:$V_{1,2,3}$] {}; 
 \node[myv2][fit=(a3),  inner xsep=0.25ex, inner ysep=0.25ex, label=above:$V_{1,3,4}$] {};

\node[myv2][fit= (y1) (y2) (y3),  inner xsep=0.5ex, inner ysep=0.5ex] {}; 

\node[myv2][fit= (y4) (y5) (y6),  inner xsep=0.5ex, inner ysep=0.5ex] {};

\node[myv2][fit= (y7) (y8) (y9),  inner xsep=0.5ex, inner ysep=0.5ex] {};

\node[myv2][fit= (y10) (y11) (y12),  inner xsep=0.5ex, inner ysep=0.5ex] {};

\node[myv2][fit= (y13) (y14) (y15),  inner xsep=0.5ex, inner ysep=0.5ex] {};

\node[myv2][fit= (y16) (y17) (y18),  inner xsep=0.5ex, inner ysep=0.5ex] {};

\node[myv2][fit= (y19) (y20) (y21),  inner xsep=0.5ex, inner ysep=0.5ex] {};

\node[myv2][fit= (y22) (y23) (y24),  inner xsep=0.5ex, inner ysep=0.5ex] {};


\node[myv4][fit= (x1) (x1') (x2) (x2') (x3) (x3') (x4) (x4'),  inner xsep=0.5ex, inner ysep=0.5ex] {};

 \draw [dashed] (a1) -- (a2);
 \draw [dashed] (a1) to[bend left=8] (a3);
 \draw [dashed] (a2) -- (a3);
 
 \draw [dashed] (a1) -- (x3);
 \draw [dashed] (a1) -- (x4');
 
  \draw [dashed] (a2) -- (x1);
 \draw [dashed] (a2) -- (x4');
 
  \draw [dashed] (a3) -- (x1);
 \draw [dashed] (a3) -- (x2');

  
  \draw [dashed] (x1) to[bend right=50] (y2);
  \draw [dashed] (x1) to[bend right=53] (y3);
  \draw [dashed] (x1) -- (y4);
  \draw [dashed] (x1) -- (y5);
  \draw [dashed] (x1) -- (y6);
  \draw [dashed] (x1') -- (y1);
  \draw [dashed] (x1') -- (y2);
  \draw [dashed] (x1') -- (y3);
  \draw [dashed] (x1') to[bend left=50] (y5);
  \draw [dashed] (x1') to[bend left=53] (y6);
  
   \draw [dashed] (x2) to[bend right=50] (y8);
  \draw [dashed] (x2) to[bend right=53] (y9);
  \draw [dashed] (x2) -- (y10);
  \draw [dashed] (x2) -- (y11);
  \draw [dashed] (x2) -- (y12);
  \draw [dashed] (x2') -- (y7);
  \draw [dashed] (x2') -- (y8);
  \draw [dashed] (x2') -- (y9);
  \draw [dashed] (x2') to[bend left=50] (y11);
  \draw [dashed] (x2') to[bend left=53] (y12);
  
   \draw [dashed] (x3) to[bend right=50] (y14);
  \draw [dashed] (x3) to[bend right=53] (y15);
  \draw [dashed] (x3) -- (y16);
  \draw [dashed] (x3) -- (y17);
  \draw [dashed] (x3) -- (y18);
  \draw [dashed] (x3') -- (y13);
  \draw [dashed] (x3') -- (y14);
  \draw [dashed] (x3') -- (y15);
  \draw [dashed] (x3') to[bend left=50] (y17);
  \draw [dashed] (x3') to[bend left=53] (y18);
  
  \draw [dashed] (x4) to[bend right=50] (y20);
  \draw [dashed] (x4) to[bend right=53] (y21);
  \draw [dashed] (x4) -- (y22);
  \draw [dashed] (x4) -- (y23);
  \draw [dashed] (x4) -- (y24);
  \draw [dashed] (x4') -- (y19);
  \draw [dashed] (x4') -- (y20);
  \draw [dashed] (x4') -- (y21);
  \draw [dashed] (x4') to[bend left=50] (y23);
  \draw [dashed] (x4') to[bend left=53] (y24);
  
  \draw [dashed] (y1) to[bend right=50] (y3);
  \draw [dashed] (y1) -- (y4);
  \draw [dashed] (y1) -- (y5);
  \draw [dashed] (y1) -- (y6);
  \draw [dashed] (y2) -- (y4);
  \draw [dashed] (y2) -- (y5);
  \draw [dashed] (y2) -- (y6);
  \draw [dashed] (y3) -- (y4);
  \draw [dashed] (y3) -- (y5);
  \draw [dashed] (y3) -- (y6);
  \draw [dashed] (y4) to[bend left=50] (y6);
  
   \draw [dashed] (y7) to[bend right=50] (y9);
  \draw [dashed] (y7) -- (y10);
  \draw [dashed] (y7) -- (y11);
  \draw [dashed] (y7) -- (y12);
  \draw [dashed] (y8) -- (y10);
  \draw [dashed] (y8) -- (y11);
  \draw [dashed] (y8) -- (y12);
  \draw [dashed] (y9) -- (y10);
  \draw [dashed] (y9) -- (y11);
  \draw [dashed] (y9) -- (y12);
  \draw [dashed] (y10) to[bend left=50] (y12);
  
  \draw [dashed] (y13) to[bend right=50] (y15);
  \draw [dashed] (y13) -- (y16);
  \draw [dashed] (y13) -- (y17);
  \draw [dashed] (y13) -- (y18);
  \draw [dashed] (y14) -- (y16);
  \draw [dashed] (y14) -- (y17);
  \draw [dashed] (y14) -- (y18);
  \draw [dashed] (y15) -- (y16);
  \draw [dashed] (y15) -- (y17);
  \draw [dashed] (y15) -- (y18);
  \draw [dashed] (y16) to[bend left=50] (y18);
  
  \draw [dashed] (y19) to[bend right=50] (y21);
  \draw [dashed] (y19) -- (y22);
  \draw [dashed] (y19) -- (y23);
  \draw [dashed] (y19) -- (y24);
  \draw [dashed] (y20) -- (y22);
  \draw [dashed] (y20) -- (y23);
  \draw [dashed] (y20) -- (y24);
  \draw [dashed] (y21) -- (y22);
  \draw [dashed] (y21) -- (y23);
  \draw [dashed] (y21) -- (y24);
  \draw [dashed] (y22) to[bend left=50] (y24);


\end{tikzpicture}  
  \caption{An example of Construction \ref{constructionp7} for the formula $\phi = C_1$ 
  where $C_1=x_1\lor \overline{x_2}\lor x_3\lor \overline{x_4}$. The lines (dashed) connecting two rectangles indicate that each vertex in one rectangle is \textit{nonadjacent} to all vertices in the other rectangle. Similarly the lines (dashed) connecting a circle and a rectangle indicate that the vertex corresponding to circle is \textit{nonadjacent} to all vertices in the  rectangle. 
  If there is no line shown between two entities (rectangle/circle), then the vertices in them are all-adjacent, with an exception -- 
  all the vertices in the red rectangle (dashed) together form an independent set.}
    \label{p7fig}
  \end{figure}

\begin{construction}
\label{constructionp7}
    Let $\Phi$ be an input to the construction, where $\Phi$, with $n$ variables and $m$ clauses,  is a \FSATO\ formula.
    We construct $G$ 
    in the following way:
\begin{itemize}
    \item For each variable $X_i$ in $\Phi$, introduce two vertices - one vertex, denoted by $u_i$, 
    for the positive literal $x_i$  and one vertex, denoted by $u'_i$, for the negative literal $\overline{x_i}$. 
    For each vertex $u_i$, introduce three sets $U_{i,1},U_{i,2}$, and $U_{i,3}$. 
    Each of these sets $U_{i,s}$ 
    induces a $\overline{P_7}$. 
    The vertex $u_i$ is adjacent to each vertex in $U_{i,1}$. 
    Further, the set $U_{i,s}$ is all-adjacent to $U_{i,s+1}$, for $s=\{1,2\}$. 
    That is, for any three vertices $u\in U_{i,1},v\in U_{i,2}$, and $w\in U_{i,3}$,  a $P_3$ is induced 
    by $u,v,w$ with $v$ as the degree-2 vertex of the $P_3$. 
    Similarly, for each vertex $u_i'$, introduce three sets $U'_{i,1},U'_{i,2}$, and $U'_{i,3}$.
    Each of these sets $U'_{i,s}$ 
    induces a $\overline{P_7}$. 
    The vertex $u_i'$ is all-adjacent to $U'_{i,1}$. Further, the set $U'_{i,s}$ 
    is all-adjacent to $U'_{i,s+1}$, for $s=\{1,2\}$, i.e., 
    for any three vertices $u'\in U_{i,1}',v'\in U'_{i,2}, w'\in U'_{i,3}$, 
    a $P_3$ is induced by $u',v',w'$  with $v'$ as the degree-2 vertex. 
    Thus, the total number of vertices corresponding to a variable of $\Phi$ is 44.
    
    \item  For each clause $C_i$ of the form  $\ell_{i,1}\lor \ell_{i,2}\lor \ell_{i,3}\lor \ell_{i,4}$ in $\Phi$, 
    introduce three sets  $V_{i,1,2}$, $V_{i,2,3},$  and $V_{i,3,4}$ of seven vertices each. 
    Each of these sets induces a $\overline{P_7}$. 
    Let $V_i$ denote the union of all these three sets.
    Clearly, $V_i$ has 21 vertices.
    Let the four vertices introduced (in the previous step) for the literals $\ell_{i,1}, \ell_{i,2}, \ell_{i,3}$, and $\ell_{i,4}$ be denoted by  
    $y_{i,1},y_{i,2},y_{i,3}$, and $y_{i,4}$ respectively. 
    We note that, if $\ell_{i,1} = x_j$, then $y_{i,1} = u_j$, and 
    if $\ell_{i,1} = \overline{x_j}$, then $y_{i,1} = u'_j$. 
    Similarly, let the four  vertices introduced for the negation of these literals be denoted by  
    $z_{i,1},z_{i,2},z_{i,3}$ and $z_{i,4}$ respectively.
    We note that, if $\ell_{i,1} = x_j$, then $z_{i,1} = u'_j$, and 
    if $\ell_{i,1} = \overline{x_j}$, then $z_{i,1} = u_j$. 
    Further, every vertex in $V_{i,s,t}$ is adjacent to the vertices $y_{i,s}$ as well as $y_{i,t}$. 
    In addition to this, every vertex in $V_i$ is adjacent to all literal vertices corresponding to literals not in $C_i$.
    
    \item For all $i \neq j$, the set $V_i$ is all-adjacent to the set $V_j$.
    \item The sets $U_{i,s}$ and $U'_{i,s}$ (for $1\leq i\leq s$) are all-adjacent to $V(G)\setminus (U_i\cup U'_i\cup \{u_i, u'_i\}$).
    
   
\end{itemize} 
This completes the construction, an example of which is shown in Figure \ref{p7fig}. 
\end{construction}
For convenience, we call each set $V_i$  introduced in the construction a \textit{clause set} 
with each of them containing three subsets of vertices  $V_{i,1,2},  V_{i,2,3}$, and $V_{i,3,4}$. 
The union of all $V_i$s is denoted by $V'$.
For each variable $X_i$, the vertices $u_i$ and $u'_i$ are called \textit{literal vertices}. 
The union of all literal vertices is denoted by $U$.
Further, for each variable $X_i$, by $U_i$ we denote $U_{i,1}\cup U_{i,2}\cup U_{i,3}$
and by $U'_i$ we denote $U'_{i,1}\cup U'_{i,2}\cup U'_{i,3}$. 
We call each sets $U_i$ and $U'_i$ as \textit{hanging sets}. 
 The vertices in the hanging sets are called \textit{hanging vertices}.
Let $U^h$ denote the set of all hanging vertices.

We note that, for a clause set $V_i$, exactly one vertex each from $V_{i,s,t}$ and the literal vertices $\{y_{i,1}, y_{i,2}, y_{i,3}, y_{i,4}\}$ 
induce a $P_7$. 
Since not all vertices of a $\overline{P_7}$ can be a part of a solution, not all vertices from each set $V_{i,s,t}$ can be in a solution. 
Therefore, a solution must contain two literal vertices corresponding to each clause.
Further a solution cannot contain both $u_i$ and $u'_i$ for some $i$. In that case, these two vertices along with its hanging vertices
form a $P_7$. These observations help us to get a valid truth assignment for $\Phi$ from a solution $S$ of $G$, and a solution for $G$ from a truth assignment satisfying $\Phi$.


\begin{theorem}
\label{thm:p7}
    \SCT\ $P_7$-free graphs is NP-complete. Further, the problem cannot be solved in time $2^{o(|V(G)|)}$, assuming the ETH.
\end{theorem}

\begin{proof}
     Let $G$ be a graph obtained by Construction \ref{constructionp7} on the input $\Phi$, 
     where $\Phi$, with $n$ variables and $m$ clauses, is an instance of \FSAT\ problem. 
     The graph $G$ has $44n+21m$ vertices. Hence, the reduction is a linear reduction.
     Therefore, it is sufficient to prove that $\Phi$ is a \textit{yes}-instance of \FSAT\ problem if and only if 
     there exists a set $S\subseteq V(G)$ of vertices such that $G\oplus S$ is $P_7$-free. 

    To prove the forward direction, assume that $\Phi$ is satisfiable and let $L$ be the  set of true literals. We note that for every variable $X_i$, either $x_i$ or $\overline{x_i}$ is true but not both. 
    Let $S$ be the set of all literal vertices corresponding to the true literals, i.e., if $x_i$ is true then $u_i$ is in $S$,
    otherwise, the vertex $u'_i$ is in $S$. Our claim is that $G\oplus S$ is $P_7$-free. For a contradiction, assume that $G\oplus S$ contains $P_7$ induced by $W\subseteq V(G)$.
    
    Claim 1: Let $A$ be a set of seven vertices in $G$ such that $G[A]$ is isomorphic to a $\overline{P_7}$.
    Further, assume that $A$ is a module in $G$. Then $|W\cap A|\leq 1$.
    
    Proof of Claim 1: Since $\overline{P_7}$ is not isomorphic to $P_7$, $W\setminus A$ is not an empty set.
    Since $A$ is a module and $P_7$ is a connected graph, there is a vertex, say $v$, in $W\setminus A$
    which is adjacent to all vertices in $A$. If $W\cap A$ contains at least three vertices,
    then three from those vertices along with $v$ form a subgraph (not necessarily induced) isomorphic to $K_{1,3}$, 
    which is a contradiction. Therefore $|W\cap A|\leq 2$. 
    Assume that $W\cap A$ has two vertices.
    Then we get a contradiction as $A$ is a module and there is no pair of vertices in a $P_7$
    which has same neighborhood.
    
    Claim 2: Let $V'$ be the set of all clause vertices. 
    Then $V'\cap W$ induces an empty graph of at most three vertices, or $K_2$, or $P_3$ in $G\oplus S$.
    
    Proof of Claim 2: Assume that $V'\cap W$ induces an empty graph in $G\oplus S$. 
    By Claim 1, $W$ contains only at most one vertex from $V_{i,s,t}$. 
    Then $|V'\cap W|$ contains at most three vertices as a set $V_i$ is all-adjacent to a set $V_j$, for $i\neq j$. 
    Now, assume that $V'\cap W$ does not induce an empty graph, i.e., $V'\cap W$ induces a graph with at least one edge. 
    By Claim 1, $W$ contains only at most one vertex from $V_{i,s,t}$. 
    Therefore, the vertices in $W\cap V_i$ forms an independent set. 
    Hence, $W$ has vertices from at least two clause sets, say $V_i$ and $V_j$. 
    Since $P_7$ is triangle-free, $W$ cannot have vertices from more than three clause sets -- note that two clause sets are all-adjacent. 
    Therefore, $W$ contains vertices from exactly two clause sets $V_i$ and $V_j$. 
    If at least one of them, say $V_i$ contains three vertices in $W$, then there is a claw as a subgraph (not necessarily induced) 
    in the graph induced by $W$, which is a contradiction as there is no claw as a subgraph in $P_7$. 
    If there are two vertices from each of $V_i$ and $V_j$ in $W$, then there is a $C_4$ as a subgraph 
    (not necessarily induced) in the graph induced by $W$, which is a contradiction as there is no $C_4$ 
    as a subgraph in $P_7$. Then there are only two cases: 
    (i) both the clause sets contain exactly one vertex each from $W$;
    (ii) one of the clause set, say $V_i$, contains two vertices and the other clause set $V_j$ contains one vertex from $W$.
    In case (i), $V'\cap W$ induces $K_2$ and in case (ii), $V'\cap W$ induces $P_3$.

    Claim 3: The set $U\cap W$ induces (in $G\oplus S$) a graph with at most four vertices which is isomorphic to an empty graph or a $P_2\cup aK_1$ (for some $1\leq a\leq 2$). 
    
    Proof of Claim 3: The set  $U$ induces a $K_n\cup nK_1$ in $G\oplus S$, where
    the $K_n$ is formed by the vertices corresponding to true literals.
    If $U\cap W$ induces an empty graph then it can have only at most four vertices as a $P_7$ does not contain an independent set of size five.
    If $U\cap W$ does not induce an empty graph, then it contains exactly two vertices from $K_n$ (if there are three, then there is a triangle).
    Therefore, $U\cap W$ induces a graph isomorphic to $P_2\cup aK_1$ for some $0\leq a\leq 2$ -- note that there is no subgraph isomorphic to $P_2\cup 3K_1$
    in $P_7$.

    Claim 4: $U^h\cap W = \emptyset$. It is sufficient to prove that 
    $W\cap (U_i\cap U'_i) = \emptyset$, for some $1\leq i\leq n$.
    By Claim 1, $|U_{i,s}\cap W|\leq 1$ and $|U'_{i,s}\cap W|\leq 1$, for $1\leq s\leq 3$.
    Since $W\cap (U_i\cup U'_i\cup \{u_i,u'_i\})$ induces an induced subgraph of $2P_4$, $W$ has 
    at least one vertex, say $v$, from $V(G)\setminus (U_i\cup U'_i\cup \{u_i,u'_i\})$. 
    Then, if $U_i\cup U'_i$ has
    at least three vertices in $W$, then there is claw as a subgraph (not necessarily induced, 
    formed by $v$ and vertices in $W\cap (U_i\cup U'_i)$)
    in the graph induced by $W$, which is a contradiction.
    If $U_i\cup U'_i$ has two vertices in $W$, then $W$ has at least three vertices from $V(G)\setminus (U_i\cup U'_i\cup \{u_i,u'_i\})$.
    Then there is a $K_{1,3}$ as a subgraph in the graph induced by $W$, which is a contradiction.
    If $W$ has only one vertex in $U_i\cup U'_i$, then there are at least four vertices in $W$ from $V(G)\setminus (U_i\cup U'_i\cup \{u_i,u'_i\})$,
    which gives a similar contradiction.

    With these claims, we are ready to complete the proof of the forward direction. 
    We analyse the cases based on the structure of the graph induced by $V'\cap W$ and $U\cap W$ 
    as per Claim 2 and Claim 3 respectively. Claim 4 implies that all vertices in $W$ are from $V'\cup U$.
    
    Case 1: $|V'\cap W| \leq 2$. Then $W\cap U$ induces a graph with at least five vertices, which is a contradiction as per Claim 3.
    
    Case 2: $|V'\cap W| = 3$. Then by Claim 2, $V'\cap W$ induces either a $3K_1$ or a $P_3$.
    Case 2(i): $V'\cap W$ induces $3K_1$. This implies that $V'\cap W$ has exactly one vertex each from
    $V_{i,1,2}, V_{i,2,3}$, and $V_{i,3,4}$ for some $i$. Then $W\cap U$ induces either $4K_1$ or $P_2\cup 2K_1$.
    In both the cases, every vertex in $U\cap W$ is nonadjacent to at least one vertex in $V_i\cap W$.
    This implies that $W\cap U=\{y_{i,1}, y_{i,2}, y_{i,3}, y_{i,4}\}$. Since at least two vertices in $U\cap W$
    is in $S$, $W$ induces a graph with at least seven edges, 
    a contradiction. 
    Case 2(ii): $V'\cap W$ induces a $P_3$. 
    Then $W\cap U$
    induces either a $P_4$, or a $P_3\cup P_1$, or a $2P_2$. All these are dismissed by Claim 3. 
    
    Case 3: $|V'\cap W| \geq 4$.
    This case is dismissed by Claim 2.

    To prove the other direction, assume that $G$ is a yes-instance, and $S\subseteq V(G)$ is a solution, i.e., $G\oplus S$ is $P_7$-free. We claim that $\Phi$ is satisfiable with the truth assignment as follows:

\begin{center}
 \begin{math} X_i=\left\{\begin{array}{ll}True, & \mbox{if the vertex $u_i$ is in $S$}.\\
      False, & \mbox{otherwise}.
    \end{array}
  \right.
\end{math}
\end{center}
    
For a contradiction, assume that $\phi$ is not satisfiable.



    Since each $V_{i,s,t}$ induces a $\overline{P_7}$, there is at least one vertex in $V_{i,s,t}$ which is not in $S$.
    Similarly, for $1\leq i\leq n$, at least one vertex each from $U_{i,s}$ (for $1\leq s\leq 3$) and at least one vertex each from $U'_{i,s}$ are not in $S$
    as these sets induce a $\overline{P_7}$ in $G$. 
    We claim that both $u_i$ and $u'_i$ cannot be in $S$ together. 
    If this happen, then these two vertices along with exactly one vertex each from $S \setminus U_{i,s}$ (for $1\leq s\leq 3$) 
    and exactly one vertex each from $S \setminus U'_{i,s}$ (for $1\leq s\leq 3$) induces a $P_8$, which is a contradiction.
    Therefore, if a literal vertex is in $S$, then the corresponding literal is assigned True by $\Phi$.
    Assume that only at most one literal in $C_i$ is true. This implies that only at most one vertex 
    in $\{y_{i,1}, y_{i,2}, y_{i,3}, y_{i,4}\}$ is in $S$. This implies that these four vertices along with 
    exactly one vetex each from $S\setminus V_{i,1,2}$, $S\setminus V_{i,2,3}$, and $S\setminus V_{i,3,4}$ induce a $P_7$ in $G\oplus S$,
    which is a contradiction.
    \end{proof}


The case when $H=P_8$ is handled in a very similar manner. In the new construction, for every clause $C_i$, there is an extra set $V_{i,1}$ which induces
a $\overline{P_8}$ and is all-adjacent to the literal vertex $y_{i,1}$. 

\begin{figure}[ht]
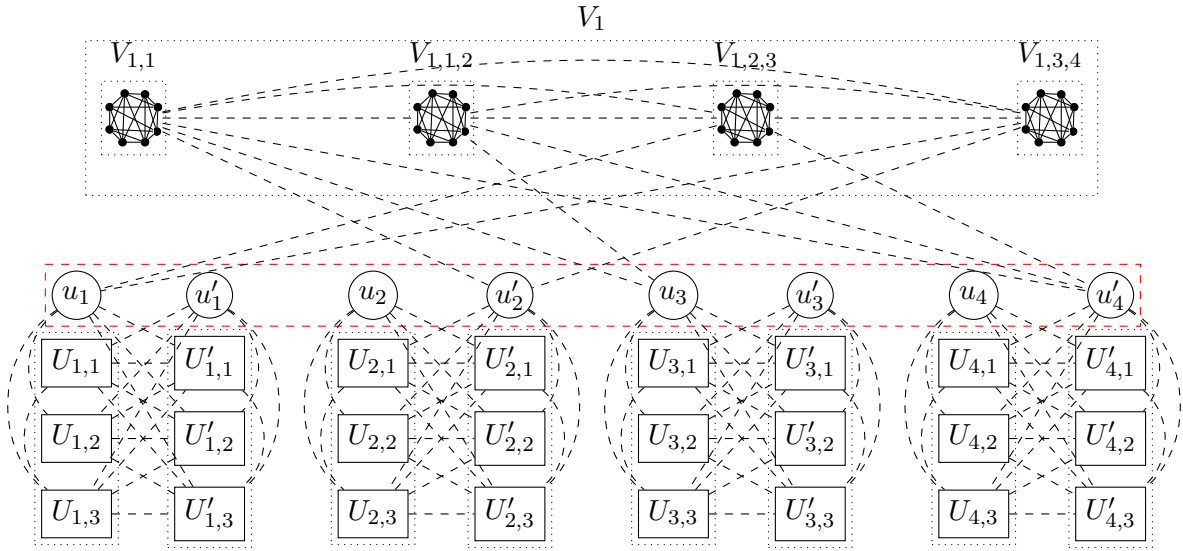

  \centering
    \centering
    \begin{tikzpicture}  [myv/.style={circle, draw, inner sep=2pt},myv1/.style={rectangle, draw},myv2/.style={rectangle, draw,dotted},myv3/.style={circle, draw, inner sep=0.5pt},my/.style={\input{figs/hard/c},draw}, myv4/.style={rectangle, draw,dashed,color=red}] 
 
\node[inner sep=0pt] (a1) at (4,5)
    {\input{figs/hard/p8c}};
\node[inner sep=0pt] (a2) at (8,5) {\input{figs/hard/p8c}};
\node[inner sep=0pt] (a3) at (12,5)
    {\input{figs/hard/p8c}};
    
\node[inner sep=0pt] (a4) at (-0.05,5)
    {\input{figs/hard/p8c}};

   \node[myv] (x1) at (-0.8,2.65) {$u_{1}$};
   \node[myv3] (x1') at (0.95,2.65) {$u'_{1}$};
   \node[myv] (x2) at (3.10,2.65) {$u_{2}$};
   \node[myv3] (x2') at (4.90,2.65) {$u'_{2}$};
   \node[myv] (x3) at (7.05,2.65) {$u_{3}$};
   \node[myv3] (x3') at (8.85,2.65) {$u'_{3}$};
   \node[myv] (x4) at (11,2.65) {$u_{4}$};
   \node[myv3] (x4') at (12.8,2.65) {$u'_{4}$};

    \node[myv1]  (y1) at (-0.8,1.75) {$U_{1,1}$};
    \node[myv1]  (y2) at (-0.8,0.75) {$U_{1,2}$};
    \node[myv1]  (y3) at (-0.8,-0.25) {$U_{1,3}$};
    \node[myv1]  (y4) at (0.95,1.75) {$U'_{1,1}$};
    \node[myv1]  (y5) at (0.95,0.75) {$U'_{1,2}$};
    \node[myv1]  (y6) at (0.95,-0.25) {$U'_{1,3}$};
    
    \node[myv1]  (y7) at (3.10,1.75) {$U_{2,1}$};
    \node[myv1]  (y8) at (3.10,0.75) {$U_{2,2}$};
    \node[myv1]  (y9) at (3.10,-0.25) {$U_{2,3}$};
    \node[myv1]  (y10) at (4.90,1.75) {$U'_{2,1}$};
    \node[myv1]  (y11) at (4.90,0.75) {$U'_{2,2}$};
    \node[myv1]  (y12) at (4.90,-0.25) {$U'_{2,3}$};
    
    \node[myv1]  (y13) at (7.05,1.75) {$U_{3,1}$};
    \node[myv1]  (y14) at (7.05,0.75) {$U_{3,2}$};
    \node[myv1]  (y15) at (7.05,-0.25) {$U_{3,3}$};
     \node[myv1]  (y16) at (8.85,1.75) {$U'_{3,1}$};
    \node[myv1]  (y17) at (8.85,0.75) {$U'_{3,2}$};
    \node[myv1]  (y18) at (8.85,-0.25) {$U'_{3,3}$};
    
     \node[myv1]  (y19) at (11,1.75) {$U_{4,1}$};
    \node[myv1]  (y20) at (11,0.75) {$U_{4,2}$};
    \node[myv1]  (y21) at (11,-0.25) {$U_{4,3}$};
     \node[myv1]  (y22) at (12.8,1.75) {$U'_{4,1}$};
    \node[myv1]  (y23) at (12.8,0.75) {$U'_{4,2}$};
    \node[myv1]  (y24) at (12.8,-0.25) {$U'_{4,3}$};

  
 \node[myv2][fit=(a4) (a3),  inner xsep=1.5ex, inner ysep=3.5ex, label=above:$V_1$] {}; 
 
 
 
 \node[myv2][fit=(a1),  inner xsep=0.25ex, inner ysep=0.25ex, label=above:$V_{1,1,2}$] {}; 
 \node[myv2][fit=(a2),  inner xsep=0.25ex, inner ysep=0.25ex, label=above:$V_{1,2,3}$] {}; 
 \node[myv2][fit=(a3),  inner xsep=0.25ex, inner ysep=0.25ex, label=above:$V_{1,3,4}$] {}; 
 
  \node[myv2][fit=(a4),  inner xsep=0.25ex, inner ysep=0.25ex, label=above:$V_{1,1}$] {};

\node[myv2][fit= (y1) (y2) (y3),  inner xsep=0.5ex, inner ysep=0.5ex] {}; 

\node[myv2][fit= (y4) (y5) (y6),  inner xsep=0.5ex, inner ysep=0.5ex] {};

\node[myv2][fit= (y7) (y8) (y9),  inner xsep=0.5ex, inner ysep=0.5ex] {};

\node[myv2][fit= (y10) (y11) (y12),  inner xsep=0.5ex, inner ysep=0.5ex] {};

\node[myv2][fit= (y13) (y14) (y15),  inner xsep=0.5ex, inner ysep=0.5ex] {};

\node[myv2][fit= (y16) (y17) (y18),  inner xsep=0.5ex, inner ysep=0.5ex] {};

\node[myv2][fit= (y19) (y20) (y21),  inner xsep=0.5ex, inner ysep=0.5ex] {};

\node[myv2][fit= (y22) (y23) (y24),  inner xsep=0.5ex, inner ysep=0.5ex] {};


\node[myv4][fit= (x1) (x1') (x2) (x2') (x3) (x3') (x4) (x4'),  inner xsep=0.5ex, inner ysep=0.5ex] {};

 \draw [dashed] (a1) -- (a2);
 \draw [dashed] (a1) to[bend left=10] (a3);
 \draw [dashed] (a2) -- (a3);
 \draw [dashed] (a4) -- (a1);
 \draw [dashed] (a4) to[bend left=10] (a2);
 \draw [dashed] (a4) to[bend left=12] (a3);
 
 \draw [dashed] (a1) -- (x3);
 \draw [dashed] (a1) -- (x4');
 
  \draw [dashed] (a2) -- (x1);
 \draw [dashed] (a2) -- (x4');
 
  \draw [dashed] (a3) -- (x1);
 \draw [dashed] (a3) -- (x2');

\draw [dashed] (a4) -- (x2');
\draw [dashed] (a4) -- (x3);
\draw [dashed] (a4) -- (x4');

  
  \draw [dashed] (x1) to[bend right=50] (y2);
  \draw [dashed] (x1) to[bend right=53] (y3);
  \draw [dashed] (x1) -- (y4);
  \draw [dashed] (x1) -- (y5);
  \draw [dashed] (x1) -- (y6);
  \draw [dashed] (x1') -- (y1);
  \draw [dashed] (x1') -- (y2);
  \draw [dashed] (x1') -- (y3);
  \draw [dashed] (x1') to[bend left=50] (y5);
  \draw [dashed] (x1') to[bend left=53] (y6);
  
   \draw [dashed] (x2) to[bend right=50] (y8);
  \draw [dashed] (x2) to[bend right=53] (y9);
  \draw [dashed] (x2) -- (y10);
  \draw [dashed] (x2) -- (y11);
  \draw [dashed] (x2) -- (y12);
  \draw [dashed] (x2') -- (y7);
  \draw [dashed] (x2') -- (y8);
  \draw [dashed] (x2') -- (y9);
  \draw [dashed] (x2') to[bend left=50] (y11);
  \draw [dashed] (x2') to[bend left=53] (y12);
  
   \draw [dashed] (x3) to[bend right=50] (y14);
  \draw [dashed] (x3) to[bend right=53] (y15);
  \draw [dashed] (x3) -- (y16);
  \draw [dashed] (x3) -- (y17);
  \draw [dashed] (x3) -- (y18);
  \draw [dashed] (x3') -- (y13);
  \draw [dashed] (x3') -- (y14);
  \draw [dashed] (x3') -- (y15);
  \draw [dashed] (x3') to[bend left=50] (y17);
  \draw [dashed] (x3') to[bend left=53] (y18);
  
  \draw [dashed] (x4) to[bend right=50] (y20);
  \draw [dashed] (x4) to[bend right=53] (y21);
  \draw [dashed] (x4) -- (y22);
  \draw [dashed] (x4) -- (y23);
  \draw [dashed] (x4) -- (y24);
  \draw [dashed] (x4') -- (y19);
  \draw [dashed] (x4') -- (y20);
  \draw [dashed] (x4') -- (y21);
  \draw [dashed] (x4') to[bend left=50] (y23);
  \draw [dashed] (x4') to[bend left=53] (y24);
  
  \draw [dashed] (y1) to[bend right=50] (y3);
  \draw [dashed] (y1) -- (y4);
  \draw [dashed] (y1) -- (y5);
  \draw [dashed] (y1) -- (y6);
  \draw [dashed] (y2) -- (y4);
  \draw [dashed] (y2) -- (y5);
  \draw [dashed] (y2) -- (y6);
  \draw [dashed] (y3) -- (y4);
  \draw [dashed] (y3) -- (y5);
  \draw [dashed] (y3) -- (y6);
  \draw [dashed] (y4) to[bend left=50] (y6);
  
   \draw [dashed] (y7) to[bend right=50] (y9);
  \draw [dashed] (y7) -- (y10);
  \draw [dashed] (y7) -- (y11);
  \draw [dashed] (y7) -- (y12);
  \draw [dashed] (y8) -- (y10);
  \draw [dashed] (y8) -- (y11);
  \draw [dashed] (y8) -- (y12);
  \draw [dashed] (y9) -- (y10);
  \draw [dashed] (y9) -- (y11);
  \draw [dashed] (y9) -- (y12);
  \draw [dashed] (y10) to[bend left=50] (y12);
  
  \draw [dashed] (y13) to[bend right=50] (y15);
  \draw [dashed] (y13) -- (y16);
  \draw [dashed] (y13) -- (y17);
  \draw [dashed] (y13) -- (y18);
  \draw [dashed] (y14) -- (y16);
  \draw [dashed] (y14) -- (y17);
  \draw [dashed] (y14) -- (y18);
  \draw [dashed] (y15) -- (y16);
  \draw [dashed] (y15) -- (y17);
  \draw [dashed] (y15) -- (y18);
  \draw [dashed] (y16) to[bend left=50] (y18);
  
  \draw [dashed] (y19) to[bend right=50] (y21);
  \draw [dashed] (y19) -- (y22);
  \draw [dashed] (y19) -- (y23);
  \draw [dashed] (y19) -- (y24);
  \draw [dashed] (y20) -- (y22);
  \draw [dashed] (y20) -- (y23);
  \draw [dashed] (y20) -- (y24);
  \draw [dashed] (y21) -- (y22);
  \draw [dashed] (y21) -- (y23);
  \draw [dashed] (y21) -- (y24);
  \draw [dashed] (y22) to[bend left=50] (y24);


\end{tikzpicture}  
  \caption{An example of Construction \ref{constructionp8} for the formula $\phi = C_1$ 
  where $C_1=x_1\lor \overline{x_2}\lor x_3\lor \overline{x_4}$. The lines (dashed) connecting two rectangles indicate that each vertex in one rectangle is \textit{nonadjacent} to all vertices in the other rectangle. Similarly the lines (dashed) connecting a circle and a rectangle indicate that the vertex corresponding to circle is \textit{nonadjacent} to all vertices in the  rectangle. 
  If there is no line shown between two entities (rectangle/circle), then the vertices in them are all-adjacent, with an exception -- 
  all the vertices in the red rectangle (dashed) together form an independent set.}
    \label{p8fig}
  \end{figure}

\begin{construction}
\label{constructionp8}
    Let $\Phi$ be an input to the construction, where $\Phi$, with $n$ variables and $m$ clauses,  is a \FSATO\ formula.
    We construct $G$ in the following way:
\begin{itemize}
    \item For each variable $X_i$ in $\Phi$, introduce two vertices - one vertex, denoted by $u_i$, 
    for the positive literal $x_i$  and one vertex, denoted by $u'_i$, for the negative literal $\overline{x_i}$. 
    For each vertex $u_i$, introduce three sets $U_{i,1},U_{i,2}$ and $U_{i,3}$. 
    Each of these sets $U_{i,s}$
    induces a $\overline{P_8}$. 
    The vertex $u_i$ is adjacent to each vertex in $U_{i,1}$. 
    Further, the set $U_{i,s}$ is all-adjacent to $U_{i,s+1}$, for $s=\{1,2\}$. 
    That is, for any three vertices $u\in U_{i,1},v\in U_{i,2}$ and $w\in U_{i,3}$,  a $P_3$ is induced 
    by $u,v,w$ with $v$ as the degree-2 vertex of the $P_3$. 
    Similarly, for each vertex $u_i'$, introduce three sets $U'_{i,1},U'_{i,2}$ and $U'_{i,3}$.
    Each of these sets $U'_{i,s}$ 
    induces a $\overline{P_8}$. 
    The vertex $u_i'$ is all-adjacent $U'_{i,1}$. The set $U'_{i,s}$ 
    is all-adjacent to $U'_{i,s+1}$, for $s=\{1,2\}$, i.e., 
    for any three vertices $u'\in U_{i,1}',v'\in U'_{i,2}, w'\in U'_{i,3}$, 
    a $P_3$ is induced by $u',v',w'$  with $v'$ as the degree-2 vertex. 
    Thus, the total number of vertices corresponding to a variable of $\Phi$ is 50.
    
    \item  For each clause $C_i$ of the form  $\ell_{i,1}\lor \ell_{i,2}\lor \ell_{i,3}\lor \ell_{i,4}$ in $\Phi$, introduce four sets  $V_{i,1},V_{i,1,2}$, $V_{i,2,3},$  and $V_{i,3,4}$ of eight vertices each. 
    Each of these sets induces a $\overline{P_8}$. 
    Let $V_i$ denote the union of all these four sets.
    Clearly, there are $32$ vertices in $V_i$.
    Let the four vertices introduced (in the previous step) for the literals $\ell_{i,1}, \ell_{i,2}, \ell_{i,3}$, and $\ell_{i,4}$ be denoted by  
    $y_{i,1},y_{i,2},y_{i,3}$, and $y_{i,4}$ respectively. 
    We note that, if $\ell_{i,1} = x_j$, then $y_{i,1} = u_j$, and 
    if $\ell_{i,1} = \overline{x_j}$, then $y_{i,1} = u'_j$. 
    Similarly, let the four  vertices introduced for the negation of these literals be denoted by  
    $z_{i,1},z_{i,2},z_{i,3}$ and $z_{i,4}$ respectively.
    We note that, if $\ell_{i,1} = x_j$, then $z_{i,1} = u'_j$, and 
    if $\ell_{i,1} = \overline{x_j}$, then $z_{i,1} = u_j$. 
    Further, every vertex in $V_{i,s,t}$ is adjacent to the vertices $y_{i,s}$ as well as $y_{i,t}$. Similarly, every vertex in $V_{i,1}$ is adjacent to the vertex $y_{i,1}$.
    In addition to this, every vertex in $V_i$ is adjacent to all literal vertices corresponding to literals not in $C_i$.
    
    \item For all $i \neq j$, the set $V_i$ is all-adjacent to the set $V_j$.
    \item The sets $U_{i,s}$ and $U'_{i,s}$ (for $1\leq i\leq s$) are all-adjacent to $V(G)\setminus (U_i\cup U'_i\cup \{u_i, u'_i\}$).
    
   
\end{itemize} 
This completes the construction (refer Figure \ref{p8fig}).
\end{construction}
For convenience, we call each set $V_i$  introduced in the construction a \textit{clause set} 
with each of them containing four subsets of vertices  $V_{i,1}, V_{i,1,2},  V_{i,2,3}$, and $V_{i,3,4}$.
The union of all $V_i$s is denoted by $V'$.
For each variable $X_i$, the vertices $u_i$ and $u'_i$ are called \textit{literal vertices}. 
The union of all literal vertices is denoted by $U$.
Further, for each variable $X_i$, by $U_i$, we denote $U_{i,1}\cup U_{i,2}\cup U_{i,3}$,
and by $U'_i$, we denote $U'_{i,1}\cup U'_{i,2}\cup U'_{i,3}$. 
We call each sets $U_i$ and $U'_i$ as \textit{hanging sets}. 
 The vertices in the hanging sets are called \textit{hanging vertices}.
Let $U^h$ denote the set of all hanging vertices.

\begin{theorem}
\label{thm:p8}
    \SCT\ $P_8$-free graphs is NP-complete. Further, the problem cannot be solved in time $2^{o(|V(G)|)}$, assuming the ETH.
\end{theorem}

\begin{proof}
     Let $G$ be a graph obtained by Construction \ref{constructionp8} on the input $\Phi$, 
     where $\Phi$, with $n$ variables and $m$ clauses, is an instance of \FSAT\ problem. 
     The graph $G$ has $50n+32m$ vertices. Hence, the reduction is a linear reduction.
     Therefore, it is sufficient to prove that $\Phi$ is a \textit{yes}-instance of \FSAT\ problem if and only if 
     there exists a set $S\subseteq V(G)$ of vertices such that $G\oplus S$ is $P_8$-free. 

    To prove the forward direction, assume that $\Phi$ is satisfiable and let $L$ be the  set of true literals. 
    We note that, for every variable $X_i$, either $x_i$ or $\overline{x_i}$ is true but not both. 
    Let $S$ be the set of all literal vertices corresponding to the true literals, i.e., if $x_i$ is true then $u_i$ is in $S$,
    otherwise, the vertex $u'_i$ is in $S$. Our claim is that $G\oplus S$ is $P_8$-free. For a contradiction, assume that $G\oplus S$ contains $P_8$ induced by $W\subseteq V(G)$.
    
    Claim 1: Let $A$ be a set of eight vertices in $G$ such that $G[A]$ is isomorphic to a $\overline{P_8}$.
    Further, assume that $A$ is a module in $G$. Then $|W\cap A|\leq 1$.
    
    Proof of Claim 1: Since $\overline{P_8}$ is not isomorphic to $P_8$, $W\setminus A$ is not an empty set.
    Since $A$ is a module and $P_8$ is a connected graph, there is a vertex, say $v$, in $W\setminus A$
    which is adjacent to all vertices in $A$. If $W\cap A$ contains at least three vertices,
    then three from those vertices along with $v$ form a subgraph (not necessarily induced) isomorphic to $K_{1,3}$, 
    which is a contradiction. Therefore $|W\cap A|\leq 2$. 
    Assume that $W\cap A$ has two vertices.
    Then we get a contradiction as $A$ is a module and there is no pair of vertices in a $P_8$
    which has same neighborhood.
    
    Claim 2: 
    The set $V'\cap W$ induces an empty graph of at most four vertices, or a $K_2$, or a $P_3$ in $G\oplus S$.
    
    Proof of Claim 2: Assume that $V'\cap W$ induces an empty graph in $G\oplus S$. 
    By Claim 1, $W$ contains only at most one vertex from each of $V_{i,1}$ and $V_{i,s,t}$. 
    Then $|V'\cap W|$ contains at most four vertices as a set $V_i$ is all-adjacent to a set $V_j$, for $i\neq j$. 
    Now, assume that $V'\cap W$ does not induce an empty graph. 
    This implies that $V'\cap W$ induces a graph with at least one edge. 
    By Claim 1, $W$ contains only at most one vertex from  each of $V_{i,1}$ and $V_{i,s,t}$. 
    Therefore, the vertices in $W\cap V_i$ forms an independent set. 
    Hence, $W$ has vertices from at least two clause sets, say $V_i$ and $V_j$. 
    Since $P_8$ is triangle-free, $W$ cannot have vertices from more than three clause sets -- note that two clause sets are all-adjacent. 
    Therefore, $W$ contains vertices from exactly two clause sets $V_i$ and $V_j$. 
    If at least one of them, say $V_i$ contains three vertices in $W$, then there is a claw as a subgraph (not necessarily induced) 
    in the graph induced by $W$, which is a contradiction as there is no claw as a subgraph in $P_8$. 
    If there are two vertices from each of $V_i$ and $V_j$ in $W$, then there is a $C_4$ as a subgraph 
    (not necessarily induced) in the graph induced by $W$, which is a contradiction as there is no $C_4$ 
    as a subgraph in $P_8$. Then there are only two cases: 
    (i) both the clause sets contain exactly one vertex each from $W$;
    (ii) one of the clause set, say $V_i$, contains two vertices and the other clause set $V_j$ contains one vertex from $W$.
    In case (i), $V'\cap W$ induces $K_2$ and in case (ii), $V'\cap W$ induces $P_3$.

    Claim 3: 
    The set $U\cap W$ induces (in $G\oplus S$) either an empty graph of at most four vertices
    or a $P_2\cup aK_1$ for some $0\leq a \leq 3$. 
    
    Proof of Claim 3: The set  $U$ induces a $K_n\cup nK_1$ in $G\oplus S$, where
    the $K_n$ is formed by the vertices corresponding to true literals.
    If $U\cap W$ induces an empty graph, then it must be of size at most four as there is no independent set of size five in a $P_8$.
    Assume that $W\cap U$ induces a graph with at least one edge.
    Then $U\cap W$ induces a graph isomorphic to $P_2\cup aK_1$ for some $0\leq a\leq 3$ -- note that there are 
    only at most two vertices in $W$ from $K_n$ and $P_8$ does not have an induced subgraph isomorphic to $P_2\cup 4K_1$.

    Claim 4: $U^h\cap W = \emptyset$. It is sufficient to prove that 
    $W\cap (U_i\cap U'_i) = \emptyset$, for some $1\leq i\leq n$.
    By Claim 1, $|U_{i,s}\cap W|\leq 1$ and $|U'_{i,s}\cap W|\leq 1$, for $1\leq s\leq 3$.
    Since $W\cap (U_i\cup U'_i\cup \{u_i,u'_i\})$ induces an induced subgraph of $2P_4$, $W$ has 
    at least one vertex, say $v$, from $V(G)\setminus (U_i\cup U'_i\cup \{u_i,u'_i\})$. 
    Then, if $U_i\cup U'_i$ has
    at least three vertices in $W$, then there is claw as a subgraph (formed by $v$ and vertices in $W\cap (U_i\cup U'_i)$)
    in the graph induced by $W$, which is a contradiction.
    If $U_i\cup U'_i$ has two vertices in $W$, then $W$ has at least four vertices from $V(G)\setminus (U_i\cup U'_i\cup \{u_i,u'_i\})$.
    Then there is a $K_{1,3}$ as a subgraph in the graph induced by $W$, which is a contradiction.
    If $W$ has only one vertex in $U_i\cup U'_i$, then there are at least five vertices in $W$ from $V(G)\setminus (U_i\cup U'_i\cup \{u_i,u'_i\})$,
    which gives a similar contradiction.

    With these claims, we are ready to complete the proof of the forward direction. 
    We analyse the cases based on the structure of the graph induced by $V'\cap W$ and $U\cap W$ 
    as per Claim 2 and Claim 3 respectively. Claim 4 implies that all vertices in $W$ are from $V'\cup U$.
    
    Case 1: $|V'\cap W| \leq 2$. Then $W\cap U$ induces a graph with at least six vertices, which is a contradiction as per Claim 3.
    
    Case 2: $|V'\cap W| = 3$. Then by Claim 2, $V'\cap W$ induces either a $3K_1$ or a $P_3$.
    Case 2(i): $V'\cap W$ induces $3K_1$. This implies that $V'\cap W$ has exactly one vertex each from three of the four sets
    $V_{i,1},V_{i,1,2}, V_{i,2,3}$, and $V_{i,3,4}$ for some $i$. Then $W\cap U$ induces a $P_2\cup 3K_1$ (as per Claim 3).
    In this case, every vertex in $U\cap W$ is nonadjacent to at least one vertex in $V_i\cap W$.
    This gives a contradiction as there are only four vertices $y_{i,1}, y_{i,2}, y_{i,3}$, and $y_{i,4}$
    nonadjacent to at least one vertex in $V'\cap W$ (recall that every vertex in $V_i$ is adjacent to all literal
    vertices corresponding to literals not in $C_i$).
    Case 2(ii): $V'\cap W$ induces a $P_3$. 
    Then $W\cap U$
    induces either a $P_5$, or a $P_4\cup K_1$, or a $P_3\cup P_2$. 
    All these are dismissed by Claim 3. 
    
    Case 3: $|V'\cap W| = 4$.
    Then by Claim 2, $V'\cap W$ induces a $4K_1$.
    This implies that $V'\cap W$ has exactly one vertex each from
    $V_{i,1}, V_{i,1,2}, V_{i,2,3}$, and $V_{i,3,4}$ for some $1\leq i\leq n$. 
    Then $W\cap U$ induces a $4K_1$. Then every vertex in $U\cap W$ is nonadjacent to at least one vertex in $V_i\cap W$.
    This implies that $W\cap U = \{y_{i,1}, y_{i,2}, y_{i,3}, y_{i,4}\}$. 
    Since at least two vertices in $\{y_{i,1}, y_{i,2}, y_{i,3}, y_{i,4}\}$ is in $S$,
    $W\cap U$ induces a graph with at least one edge, which is a contradiction.
 
    To prove the other direction, assume that $G$ is a yes-instance, and $S\subseteq V(G)$ is a solution, i.e., $G\oplus S$ is $P_8$-free. We claim that $\Phi$ is satisfiable with the truth assignment for $\Phi$ as follows:

\begin{center}
 \begin{math} X_i=\left\{\begin{array}{ll}True, & \mbox{if the vertex $u_i$ is in $S$}.\\
      False, & \mbox{otherwise}.
    \end{array}
  \right.
\end{math}
\end{center}

For a contradiction, assume that $\Phi$ is not satisfiable.
    Since each set $V_{i,1}$ and $V_{i,s,t}$ induces a $\overline{P_8}$, there is at least one vertex from each of  $V_{i,1}$ and $V_{i,s,t}$ which is not in $S$.
    Similarly, for $1\leq i\leq n$, at least one vertex each from $U_{i,s}$ (for $1\leq s\leq 3$) and at least one vertex each from $U'_{i,s}$ are not in $S$
    as these sets induce a $\overline{P_8}$ in $G$. 
    We claim that both $u_i$ and $u'_i$ cannot be in $S$ together. 
    If this happen, then these two vertices along with exactly one vertex each from $S \setminus U_{i,s}$ (for $1\leq s\leq 3$) 
    and exactly one vertex each from $S \setminus U'_{i,s}$ (for $1\leq s\leq 3$) induces a $P_8$, which is a contradiction.
    Therefore, if a literal vertex is in $S$, then the corresponding literal is assigned True by the truth assignment.
    Assume that only at most one literal in $C_i$ is true. This implies that only at most one vertex 
    in $\{y_{i,1}, y_{i,2}, y_{i,3}, y_{i,4}\}$ is in $S$. This implies that these four vertices along with 
    exactly one vetex each from $S\setminus V_{i,1}$, $S\setminus V_{i,1,2}$, $S\setminus V_{i,2,3}$, and $S\setminus V_{i,3,4}$ induce a $P_8$ in $G\oplus S$,
    which is a contradiction.
    \end{proof}

Theorem~\ref{thm:pt} is a direct implication of Lemma~\ref{lem:inductive:path}, Theorem~\ref{thm:p7} and Theorem~\ref{thm:p8}.
\begin{theorem}
\label{thm:pt}
Let $t\geq 7$ be any integer. Then \SCT\ $P_t$-free graphs is NP-complete.
Further, the problem cannot be solved in time $2^{o(|V(G)|)}$, unless the ETH fails. 
\end{theorem}



\subsection{Cycles}
\label{ct-free}

Here we deal with \SCT\ $H$-free graphs when $H$ is a cycle.
This subsection is very similar to the previous two subsections, 
except that the inductive reduction is from \SCT\ $P_t$-free graphs to \SCT\ $C_{t+2}$-free graphs. 
Since we already have hardness results for $H=P_t$ for $t\geq 7$, we obtain that \SCT\ $C_t$-free
graphs is NP-complete for every $t\geq 9$. Additionally, we give a reduction when $H=C_8$ so that
we obtain hardness results for every $C_t$ for $t\geq 8$.



\begin{construction}
\label{cycleconstruct}
    Let $(G', t)$ be an input to the construction, where $G'$ is a graph and $t\geq 1$ is an integer. 
    For every vertex $u$ of $G'$, 
    introduce $(t+2)$  vertices denoted by the set $W_u$.
    Each of these sets induces  a  $\overline{C_{t+2}}$.
    Further, every vertex $u\in V(G')$ is adjacent to  every vertex in $W_u$. For $i\neq j$ every vertex in $W_i$ is adjacent to every vertex in $W_j$.
    Let the resultant graph be $G$, and let $W$ be the union of all newly introduced vertices.   
\end{construction}

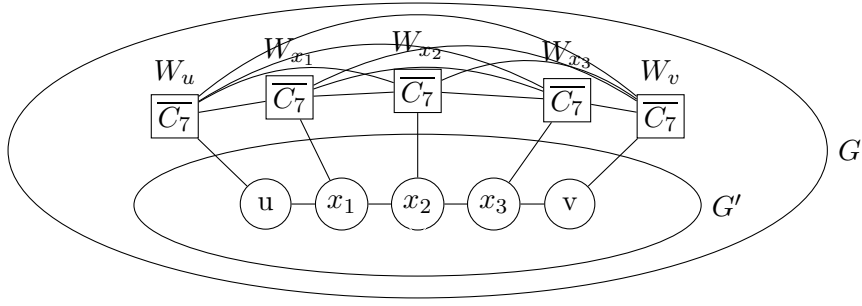
\begin{figure}[ht]
  \centering
    \centering
    \begin{tikzpicture}[myv/.style={circle, draw,color=white, inner sep=2.5pt}, myv1/.style={circle, draw, inner sep=2.5pt},myv2/.style={rectangle, draw,inner sep=16.5pt}, myv3/.style={ellipse, draw, inner sep=8.5pt},myv4/.style={rectangle, draw, inner sep=2.5pt}, myv5/.style={circle, draw, inner sep=4pt}]

  \node[myv4] [label=above: $W_{x_1}$](wa1) at (140:2.2) {$\overline{C_{7}}$};
  \node[myv4] [label=above: $W_{u}$] (wb1) at (160:3.4) {$\overline{C_{7}}$};%
  \node[myv4]  [label=above: $W_{x_3}$] (wc1) at (35:2.4) {$\overline{C_{7}}$};
  \node[myv4]  [label=above: $W_{v}$](wd1) at (20:3.4) {$\overline{C_{7}}$};
  \node[myv4] [label=above: $W_{x_2}$] (wu1) at (90:1.5) {$\overline{C_{7}}$};

 
 \node[myv3] [label=right: $G$] [fit= (a) (b) (c) (d)  (wa1) (wb1) (wc1) (wd1) (wu1)]{}; 
 
  \node[myv5] (b) at (180:2) {u};
  \node[myv1] (a) at (180:1) {$x_1$};
   \node [myv1](u) at (0,0) {$x_2$};
  \node[myv1] (c) at (180:-1) {$x_3$};
  \node[myv5] (d) at (180:-2) {v};
  \node [myv] (v) at (-90:0.25) {};
  
  \node[myv3] [label=right: $G'$] [fit=(a) (b) (c) (d) (v)]{};

  \draw (u) -- (a); 
  \draw (a) -- (b);
  \draw (u) -- (c);
  \draw (c) -- (d);
  \draw (wu1) -- (u);
  \draw (wa1) -- (a);
  \draw (wb1) -- (b);
  \draw (wc1) -- (c);
  \draw (wd1) -- (d);
 
  \draw  (wa1) -- (wb1);
  \draw  (wa1) -- (wu1);
  \draw (wc1) -- (wd1);
  \draw (wu1) -- (wc1);
\draw  (wa1) to[bend left=20] (wc1);
\draw  (wa1) to[bend left=30] (wd1);
\draw  (wb1) to[bend left=40] (wd1);
\draw  (wb1) to[bend left=25] (wu1);
\draw  (wb1) to[bend left=30] (wc1);
\draw  (wu1) to[bend left=30] (wd1);
\end{tikzpicture}
    \caption{An example of Construction \ref{cycleconstruct} for $t=5$. The lines connecting a circle and a rectangle indicate that the vertex corresponding to the circle is adjacent to all vertices in the  rectangle. Similarly, the lines connecting two rectangles indicate that each vertex in one rectangle is adjacent to all vertices in the other rectangle.} 
    \label{ct}
  \end{figure}

An example of the construction is shown in Figure~\ref{ct}.

\begin{lemma}
   \label{lem:inductive:cycle}
   Let $\mathcal{G}'$ and $\mathcal{G}$ be the classes of $P_t$-free graphs and $C_{t+2}$-free graphs respectively for any $t\geq 4$.
   If \SCGD\ is NP-complete, then so is \SCG. Further, if \SCGD\ cannot be solved in time $2^{o(|V(G)|)}$, then so is \SCG.
\end{lemma}

\begin{proof}
  Let $G'$ be an instance of \SCGD. Apply Construction~\ref{cycleconstruct} on $(G', t)$ to obtain a graph $G$. 
  Clearly, $G$ has $|V(G')|+|V(G')|\cdot (t+2)$ vertices, and hence the reduction is a linear reduction. 
  Now, it is sufficient to prove that $G'$ is a yes-instance of \SCGD\ if and only if $G$ is a yes-instance of \SCG.
  
  Let $G'$ be a yes-instance of \SCGD. Let $S'$ be a solution, i.e., $G'\oplus S'$ is $P_t$-free. 
  We claim that $G\oplus S'$ is $C_{t+2}$-free.
  Assume for a contradiction that there is a $C_{t+2}$ induced by a set $F$ of vertices in $G\oplus S'$. 
  
  Claim 1: For every vertex $u\in V(G')$, $|W_u\cap F|\leq 1$.
  
  Proof of Claim 1:
  Assume for a contradiction that $|F\cap W_u|\geq 2$.
  Since $C_{t+2}$ is not isomorphic to $\overline{C_{t+2}}$ for $t\geq 4$, we obtain that at least one vertex in
  $F$ is from outside $W_u$. Since $C_{t+2}$ is a connected graph, there is a vertex, say $v$ in $F\setminus W_u$,
  adjacent to all vertices in $F\cap W_u$.
  Then, only at most two vertices in $W_u$ are in $F$, otherwise there will be a 
  $K_{1,3}$ as a subgraph (not necessarily induced) formed by $v$ and any three vertices in $F\cap W_u$. This
  is a contradiction, as there is no $K_{1,3}$ as a subgraph in $C_{t+2}$.
  Assume that there are two vertices $a,b$ in $F\cap W_u$. Then $a$ and $b$ must be nonadjacent.
  Otherwise, $v,a,b$ forms a triangle. 
  Since $a$ and $b$ has degree two in the graph induced by $F$, there is 
  one vertex $v'\in F\setminus (W_u\cup \{v\})$ which is adjacent to $a$ but not $b$. 
  This is a contradiction as $W_u$ is a module.
  
  Claim 1 implies that if $F\cap W_u$ is nonempty, then it contains exactly one vertex. 
  Since $W_u$ is all-adjacent to $W_{u'}$ for two distinct vertices $u,u'$ in $G'$,
  $W\cap F$ has at most two vertices and if it has two vertices then they are adjacent.
  Therefore, $G'\oplus S'$ contains an induced $P_t$, which is a contradiction.
  
  

    For the other direction, assume that $G$ is a yes-instance of \SCG. 
    Then, there is a set $S\subseteq V(G)$ such that $G\oplus S$ is $C_{t+2}$-free.
    We claim that $G'\oplus S$ is $P_t$-free. 
    For a contradiction, assume that there is a set $F$ of vertices which induces a $P_t$ in $G'\oplus S$. 
    Let $u$ and $v$ be the end vertices  of the path induced by $F$.
   If $S$ contains all vertices in $W_{u}$, then it forms a $C_{t+2}$ in $G\oplus S$, which is a contradiction. 
   Therefore, there is at least one vertex $u'\in W_{u}$, which is not  in $S$. 
   Similarly, there is at least one vertex $v'\in W_{v}$  which is not  in $S$.  
   This gives a contradiction as $F\cup\{u',v'\}$ induces  $C_{t+2}$ in $G\oplus S$.
  \end{proof}  
  
Construction~\ref{constructionc8} is used for a reduction from \FSAT\ to \SCT\ $C_8$-free graphs.
\begin{construction}
\label{constructionc8}
    Let $\Phi$ be the input to the construction, where $\Phi$, with $n$ variables and $m$ clauses,  is a \FSATO\ formula. 
    We construct $G$ 
    in the following way: 
\begin{itemize}
    \item For each variable $X_i$ in $\Phi$, introduce eight vertices - four vertices, denoted by the set $U_i$, 
    for the positive literal $x_i$,  and four vertices, denoted by the set $U'_i$, for the negative literal $\overline{x_i}$ . 
    The set $U_i$ forms a clique  and so does the set $U'_i$. 
    Together they induce a  $\overline{C_8}$ --    
    note that there are two cliques of four vertices each in $\overline{C_8}$. 
    Thus, the total number of vertices corresponding to all variables of $\Phi$ is 8$\cdot$n. 
    \item 
    For each clause $C_i$ of the form  $\ell_{i,1}\wedge \ell_{i,2}\wedge \ell_{i,3}\wedge \ell_{i,4}$ in $\Phi$, 
    introduce six sets  $V_{i,1,2}, V_{i,1,3,} V_{i,1,4}, V_{i,2,3},$ $V_{i,2,4}$, and $V_{i,3,4}$ of eight vertices each. 
    Each of these sets induces a $\overline{C_8}$. 
    Let $V_i$ denote the union of all these six sets.
    Clearly, there are 48 vertices in $V_i$.
    Let the four sets of vertices introduced (in the previous step) for the literals $\ell_{i,1}, \ell_{i,2}, \ell_{i,3}$ and $\ell_{i,4}$ be denoted by  
    $Y_{i,1},Y_{i,2},Y_{i,3}$ and $Y_{i,4}$ respectively. 
    We note that, if $\ell_{i,1} = x_j$, then $Y_{i,1} = U_j$, and if $\ell_{i,1} = \overline{x_j}$, then $Y_{i,1} = U'_j$.
    Similarly, let the four sets of vertices introduced for the negation of these literals be denoted by  
    $Z_{i,1},Z_{i,2},Z_{i,3}$, and $Z_{i,4}$ respectively.
    We note that, if $\ell_{i,1} = x_j$, then $Z_{i,1} = U'_j$, and if $\ell_{i,1} = \overline{x_j}$, then $Z_{i,1} = U_j$.
    Further, every vertex in $V_{i,s,t}$ is adjacent to every vertex in $Y_{i,s}$ as well as every vertex in $Y_{i,t}$.
    \item For all $i \neq j$, every vertex in $V_i$ is adjacent to every vertex in $V_j$. 
   
\end{itemize} 
This completes the construction (refer Figure \ref{c8fig}).
\begin{figure}[ht]
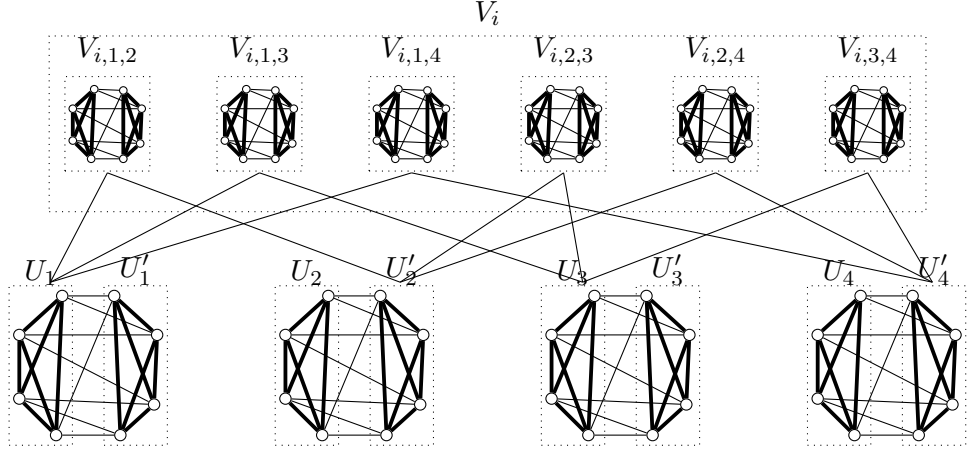

  \centering
    \centering
    \begin{tikzpicture}  [myv/.style={circle, draw},myv1/.style={rectangle, draw,dotted},my/.style={\input{figs/hard/c},draw}] 
 
\node[inner sep=0pt] (a1) at (1.25,5)
    {\input{figs/hard/c8c}};
\node[inner sep=0pt] (a2) at (3.25,5)
    {\input{figs/hard/c8c}};
\node[inner sep=0pt] (a3) at (5.25,5)
    {\input{figs/hard/c8c}};
\node[inner sep=0pt] (a4) at (7.25,5)
    {\input{figs/hard/c8c}};
\node[inner sep=0pt] (a5) at (9.25,5)
    {\input{figs/hard/c8c}};
\node[inner sep=0pt] (a6) at (11.25,5)
    {\input{figs/hard/c8c}};

    \node[inner sep=0pt] (x1) at (1,2)
    {\input{figs/hard/x1}};

     \node[inner sep=0pt] (x2) at (4.5,2)
    {\input{figs/hard/x2}};
  
 \node[inner sep=0pt] (x3) at (8,2)
    {\input{figs/hard/x3}};
  
     \node[inner sep=0pt] (x4) at (11.5,2){\input{figs/hard/x4}};

 \node[myv1][fit=(a1) (a6),  inner xsep=1.5ex, inner ysep=3.5ex, label=above:$V_i$] {};

 \node[myv1][fit=(a1),  inner xsep=0.25ex, inner ysep=0.25ex, label=above:$V_{i,1,2}$] {}; 
 \node[myv1][fit=(a2),  inner xsep=0.25ex, inner ysep=0.25ex, label=above:$V_{i,1,3}$] {}; 
 \node[myv1][fit=(a3),  inner xsep=0.25ex, inner ysep=0.25ex, label=above:$V_{i,1,4}$] {}; 
 \node[myv1][fit=(a4),  inner xsep=0.25ex, inner ysep=0.25ex, label=above:$V_{i,2,3}$] {}; 
 \node[myv1][fit=(a5),  inner xsep=0.25ex, inner ysep=0.25ex, label=above:$V_{i,2,4}$] {}; 
 \node[myv1][fit=(a6),  inner xsep=0.25ex, inner ysep=0.25ex, label=above:$V_{i,3,4}$] {};

  \draw (1.25,4.35) -- (0.5,2.9);
  \draw (1.25,4.35) -- (5.1,2.9);
  \draw (3.25,4.35) -- (0.5,2.9); 
  \draw (3.25,4.35) -- (7.5,2.9); 
  \draw (5.25,4.35) -- (0.5,2.9);  
  \draw (5.25,4.35) -- (12.1,2.9); 
  \draw (7.25,4.35) -- (5.1,2.9);  
  \draw (7.25,4.35) -- (7.5,2.9);  
  \draw (9.25,4.35) -- (5.1,2.9);  
  \draw (9.25,4.35) -- (12.1,2.9);  
  \draw (11.25,4.35) -- (7.5,2.9);  
  \draw (11.25,4.35) -- (12.1,2.9); 

\end{tikzpicture}  
    \caption{An example of Construction \ref{constructionc8} for the formula $\Phi = C_i$ where $C_i=x_1\lor \overline{x_2}\lor x_3\lor \overline{x_4}$. 
    The bold lines in each structure represents the edges in $K_4$s. The lines connecting two rectangles indicate that each vertex in one rectangle is adjacent to all vertices in the other rectangle. 
    }
    \label{c8fig}
  \end{figure}%
\end{construction}

For convenience, we call each set $V_i$  introduced in the construction a \textit{clause set} 
with each of them containing six subsets of vertices  $V_{i,1,2}, V_{i,1,3,} V_{i,1,4}, V_{i,2,3}, V_{i,2,4}$, and $V_{i,3,4}$. Similarly, for each clause, for each literal $\ell_{i,s}$ (for $1\leq s\leq 4$), the sets $Y_{i,s}$ and $Z_{i,s}$ are called \textit{literal sets}. 
The vertices in the literal sets 
are called \textit{literal vertices}. 

\begin{theorem}
\label{thm:c8}
    \SCT\ $C_8$-free graphs is NP-complete. Further, the problem cannot be solved in time $2^{o(|V(G)|)}$, assuming the ETH.
\end{theorem}

\begin{proof}
     Let $G$ be a graph obtained by Construction \ref{constructionc8} on the input $\Phi$, with $n$ variables and $m$ clauses,
     is an instance of \FSAT\ problem. 
     Since there are only $8n + 48m$ vertices in $G$, the reduction is a linear reduction. 
     Therefore, it is sufficient to prove that 
    $\Phi$ is a \textit{yes}-instance 
    of \FSAT\  if and only if there exists a set $S\subseteq V(G)$  
    such that $G\oplus S$ is $C_8$-free. 

    To prove the forward direction, assume that $\Phi$ is satisfiable and let $L$ be the  set of true literals. We note that 
    for every variable $X_i$, either $x_i$ or $\overline{x_i}$ is true but not both. 
    Let $S$ be the set of all vertices in the literal sets corresponding to the true literals, i.e,. if $x_i$ is true then all vertices in $U_i$ are in $S$,
    otherwise, all vertices in $U'_i$ are in $S$.
    Our claim is that $G\oplus S$ is $C_8$-free. 
    For a contradiction, assume that  $G\oplus S$ contains $C_8$ induced by $W\subseteq V(G)$.

    Claim 1: For every $i,s,t$ (for which $V_{i,s,t}$ is defined), $|V_{i,s,t}\cap W| \leq 1$.

    Proof of Claim 1: 
    For a contradiction, assume that there are two vertices $a,b \in V_{i,s,t}\cap W$. 
    First we prove the case in which $a$ and $b$ are nonadjacent and then we prove the case in which they are adjacent.
    Assume that $a$ and $b$ are nonadjacent.
    Since $W$ induces a $C_8$, there exists distinct nonadjacent (in $G\oplus S$) 
    vertices $a', b'\in W\setminus \{a,b\}$ such that $a'$ is a neighbor of $a$ but not of $b$,
    and $b'$ is a neighbor of $b$ but not of $a$, i.e., $\{a',a,b,b'\}$ induces a $2K_2$. 
    Since $V_{i,s,t}$ induces a $\overline{C_8}$, there is no induced $2K_2$ in $V_{i,s,t}$. Therefore, either $a'$ or $b'$ is not from $V_{i,s,t}$.
    Assume, without loss of generality, that $b'$ is not from $V_{i,s,t}$.
    Since $V_{i,s,t}$ is a module, $b'$ is adjacent to both $a$ and $b$ or noadjacent to both $a$ and $b$.
    This is a contradiction. Now, assume that $a$ and $b$ are adjacent. Let $a'$ and $b'$ be as defined above, i.e., $a'$ and $b'$ are
    nonadjacent and $a'$ is a neighbor of $a$ but not of $b$, and $b'$ is a neighbor of $b$ but not of $a$.
    Since $W$ induces a $C_8$, there exists such two nonadjacent vertices $a'$ and $b'$ in $W\setminus \{a,b\}$.
    If $a'$ and $b'$ are in $V_{i,s,t}$, we get a contradiction (recall that $a'$ and $b'$ are nonadjacent) as per the previous case. 
    Therefore, either $a'$ or $b'$ is not from $V_{i,s,t}$. Without loss of generality, assume that $b'$ is not from $V_{i,s,t}$.
    Then, either $b'$ is adjacent to both $a$ and $b$ or $b'$ is nonadjacent to both $a$ and $b$.
    This is a contradiction to the way we defined $a'$ and $b'$.
    
    Claim 2: Let $A$ be a literal set of a variable $X_i$ (i.e., $A$ is either $U_i$ or $U'_i$). Then $|A\cap W|\leq 1$.
    
    Proof of Claim 2: We note that $A$ is either a clique (if the corresponding literal is false) 
    or an independent set (if the corresponding literal is true) of four vertices in $G\oplus S$.
    Let $B$ be the other literal set of the same variable, i.e., $B$ is $U'_i$ if $A$ is $U_i$,
    and $B$ is $U_i$ if $A$ is $U'_i$. For a contradiction, assume that there are two vertices $b,c\in A\cap W$.
    First we prove the case in which $A$ corresponds to a true literal and then we prove the case in which
    $A$ corresponds to a false literal.
    Assume that $A$ corresponds to a true literal. Therefore, $A$ is an independent set and $B$ is a clique in $G\oplus S$.
    Since $b$ and $c$ are nonadjacent
    and $W$ induces a $C_8$, there exists two distinct nonadjacent vertices $b'$ and $c'$ in $W\setminus \{b,c\}$ such that 
    $b'$ is a neighbor of $b$ but not of $c$, and $c'$ is a neighbor of $c$ but not of $b$. Clearly, $b'$ and $c'$ 
    are not in $A$. Since $b'$ and $c'$
    are nonadjacent, at least one of them is not from $B$. This is a contradiction, as $A$ is a module 
    in $(G\oplus S) - B$. 
    Now, assume that $A$ corresponds to a false literal. This implies that $A$ is a clique and $B$ is an independent set.
    Then there exists two distinct nonadjacent
    vertices $b'$ and $c'$ in $W\setminus \{b,c\}$ such that $b'$ is a neighbor of $b$ but not of $c$, and $c'$ is a neighbor of $c$
    but not of $b$. If both $b'$ and $c'$ are from $B$, then we get a contradiction from the previous case 
    (note that $B$ corresponds to a true literal). Therefore, either $b'$ or $c'$ is not from $B$.
    Then, we get a contradiction as $A$ is a module in $(G\oplus S) - B$.
    
    Claim 3: Let $U$ be the union of all literal sets of vertices. Then $U\cap W$ induces 
    a graph (in $G\oplus S$) with at most five vertices and isomorphic to $P_a\cup bK_1$ for some integers $a\geq 0, b\geq 1$.

    Proof of Claim 3: To begin with, we note that, if there is an edge in the graph induced by $U\cap W$,
    then at least one of the end vertex of the edge belongs to a literal set corresponds to a true literal. 
    To see this, note that, by Claim 2, only at most one vertex from a literal set is in $W$.
    Further, there is no edge between two literal sets corresponding to two false literals. 
    Clearly, only at most two vertices from literals sets corresponding to true literals are there in $W$ as 
    every pair of literal sets corresponding to true literals are all-adjacent.
    These also imply that there is no induced $2K_2$  in the graph induced by $U\cap W$.
    Since a literal set corresponding to a true literal is not adjacent to more than one literal sets 
    corresponding to false literals, we obtain that there is no claw as an induced subgraph in the graph induced by $U\cap W$.
    Now, we analyse the cases based on the number of edges in the graph induced by $U\cap W$.
    Case (i): The set $U\cap W$ induces a graph with no edges -- then it is an empty graph with at most four 
    vertices as there is no independent set of size at least five in a $C_8$.
    Case (ii): The set $U\cap W$ induces a graph with exactly one edge. Then, clearly, it is of the form $P_2 \cup bK_1$.
    Further $b\leq 2$ as there is no induced $P_2\cup 3K_1$ in a $C_8$.
    Case (iii): The set $U\cap W$ induces a graph with exactly two edges. Since 
    there is no $2K_2$ as an induced subgraph in the graph induced by $U\cap W$, the graph induced by $U\cap W$ must be $P_3\cup bK_1$.
    Since there is no induced $P_3\cup 3K_1$ in a $C_8$, we obtain that $b\leq 2$.
    Case (iv): The set $U\cap W$ induces a graph with exactly three edges.
    Since there is no induced $2K_2$ or claw in the graph induced by $U\cap W$,
    the graph induce by it must be $P_4\cup bK_1$. Since there is no $P_4\cup 2K_1$ in a $C_8$,
    we obtain that $b\leq 1$.
    Case (v): The set $U\cap W$ induces a graph with at least four edges.
    Since there is no $2K_2$ or $K_{1,3}$ or triangle induced by $U\cap W$, it must induce a graph with an induced path of length at least four.
    By Claim 2, only at most one vertex from each literal set is in $W$.
    Since every pair of literal sets corresponding to true literals are all-adjacent, and 
    every literal set corresponding to false literals are not adjacent to any other literal set, other 
    than the one corresponding to the true literal of the same variable, $U\cap W$ induces an induced subgraph
    of a graph isomorphic to a clique where each vertex in the clique is adjacent to exactly one degree-1 vertex.
    Since such a graph does not contain an induced path of length at least four, we get a contradiction.

    Claim 4: Let $V'$ be the set of all clause vertices. Then $V'\cap W$ induces in $G\oplus S$ an empty graph 
    of at most four vertices or $K_2$ or $P_3$.
    
    Proof of Claim 4: Assume that $V'\cap W$ induces an empty graph in $G\oplus S$. Then $|V'\cap W|$ contains 
    only at most four vertices as there is no independent set of size at least five in a $C_8$.
    Now, assume that $V'\cap W$ does not induce an empty graph. This implies that $V'\cap W$ induces a graph with 
    at least one edge. By Claim 1, $W$ contains only at most one vertex from $V_{i,s,t}$. Therefore, the vertices in $W\cap V_i$
    forms an independent set. Therefore, $W$ has vertices from at least two clause sets, say $V_i$ and $V_j$. Since $C_8$ is triangle-free, 
    $W$ cannot have vertices from more than three clause sets -- note that two clause sets are all-adjacent. 
    Therefore, $W$ contains vertices from exactly two clause sets $V_i$ and $V_j$. If at least one of them, say $V_i$ contains three vertices in $W$,
    then there is a claw as a subgraph (not necessarily induced) in the graph induced by $W$, which is a contradiction as there is no claw as a subgraph in $C_8$.
    If there are two vertices from each of $V_i$ and $V_j$ in $W$, then there is a $C_4$ as a subgraph (not necessarily induced) 
    in the graph induced by $W$, which is a contradiction as there is no $C_4$ as a subgraph in $C_8$. Then there are only two cases: 
    (i) both the clause sets contain exactly one vertex each from $W$;
    (ii) one of the clause set, say $V_i$, contains two vertices and the other clause set $V_j$ contains one vertex from $W$.
    In case (i), $V'\cap W$ induces $K_2$ and in case (ii), $V'\cap W$ induces $P_3$.

    With these claims, we are ready to complete the proof of the forward direction. We analyse the cases based on the structure 
    of the graph induced by $V'\cap W$, as per Claim 4. By Claim 3, the vertices in $U$ cannot induce a $C_8$, therefore $V'\cap W$
    is nonempty.
    
    Case 1: $|V'\cap W| = 1$. Then $U\cap W$ induced $P_7$, which is a contradiction as per Claim 3.
    
    Case 2: $|V'\cap W| = 2$. Then $U\cap W$ induces either a $P_6$ (when $V'\cap W$ induces a $K_2$) 
    or a $P_5\cup K_1$ (when $V'\cap W$ induces a $2K_1$), both are contradictions as per Claim 3.
    
    Case 3: $|V'\cap W| = 3$. Then, by Claim 4, $V'\cap W$ induces either a $P_3$ or a $3K_1$. 
    If $V'\cap W$ induces a $P_3$, then $U\cap W$ induces a $P_5$, which is a contradiction as per Claim 3. 
    If $V'\cap W$ induces a $3K_1$, then $U\cap W$ induces a $P_3\cup 2K_1$ or a $2K_2\cup K_1$, the latter gives a 
    contradiction as per Claim 3. Therefore, assume that $U\cap W$ induces a $P_3\cup 2K_1$. Let $U\cap W = \{u_1,u_2,u_3,u_4,u_5\}$,
    where $\{u_1, u_2, u_3\}$ induces the $P_3$, where $u_2$ is the degree-2 vertex of the $P_3$. 
    Similarly, assume that $V'\cap W = \{v_1,v_2,v_3\}$. Clearly, these vertices are from a single clause set, say $V_i$.
    Without loss of generality, assume that $u_1$ is adjacent to $v_1$, $u_3$ is adjacent to $v_2$, $u_4$
    is adjacent to $v_1$ and $v_3$, and $u_5$ is adjacent to $v_2$ and $v_3$. 
    We note that $u_2$ and one vertex in $\{u_1,u_3\}$ are in literal sets corresponding to true literals, and $u_4$ and $u_5$
    are in literals sets corresponding to false literals (note that there are no edges between $u_2, u_4$ and $u_5$).
    We also note that $u_1, u_3, u_4,$ and $u_5$ are the vertices from literal sets corresponding to literals in the clause $C_i$.
    But, only one of them belongs to the literal sets corresponding to the true literals. This is a contradiction as at least 
    two literals are true in clause $C_i$.
    
    Case 4: $|V'\cap W| = 4$. Then, by Claim 4, $V'\cap W$ induces an empty graph of four vertices. Clearly, all of them
    are from a single clause set, say $V_i$.
    This implies that $U\cap W$ induces an empty graph of four vertices - only at most one of them can be from a literal set 
    corresponding to a true literal. This is a contradiction, as at least two of the literals in clause $C_i$ are true.

  To prove the other direction,
assume that $G$ is a yes-instance, and $S\subseteq V(G)$ is a solution, i.e., $G\oplus S$ is $C_8$-free. We claim that $\Phi$ is satisfiable with   the truth assignment for $\Phi$ as follows:
\begin{center}
 \begin{math} X_i=\left\{\begin{array}{ll}True, & \mbox{if all vertices in $U_i$ are in $S$}.\\
      False, & \mbox{otherwise}.
    \end{array}
  \right.
\end{math}
\end{center}



Claim 5: The set $S$ contains at least one literal vertex and does not contain all vertices of $V_{i,s,t}$ for any $i,s,t$ (for which $V_{i,s,t}$ is defined). 

Proof of Claim 5: Assume that all the vertices in $V_{i,s,t}$ (for any $i,s,t$) are in $S$. These vertices form a $C_{8}$ in $G\oplus S$ because each of the sets $V_{i,x,y}$ form a $\overline{C_8}$ in $G$. Thus, at least one vertex from  each $V_{i,x,y}$ for every clause set $V_i$ is not in $S$. To prove that there is at least one literal vertex in $S$, assume, for a contradiction, 
that $S$ contains vertices only from clause sets.   
Now, form a set $W$ by taking 
one vertex, which is not in $S$, from $V_{i,1,2}$, $V_{i,2,3}$, $V_{i,3,4}$, and $V_{i,1,4}$ for some $i$. 
Add exactly one vertex each to $W$ from the literal sets corresponding to the literals of the clause $C_i$. 
Now, it can be easily verified that $W$ induces a $C_8$ in $G\oplus S$, which is a contradiction.

With this claim, we are ready to complete the proof of the backward direction. 
Since $U_i\cup U'_i$ 
induces a $\overline{C_8}$ in $G$, all the vertices in $U_i\cup U'_i$ are not in $S$. This implies that, if all vertices of 
a literal set are in $S$, then that literal is assigned true. 
To prove the backward direction, we need to prove that 
for every $i$, at least two literals are true in the clause $C_i$. 
For a contradiction, assume that only at most one literal is true for a clause $C_i$.
This implies that all the vertices of only at most one literal set corresponding to the literals of the clause $C_i$
are in $S$. Now, form a set $W$ with exactly one vertex from each literal set corresponding to the literals of the clause $C_i$.
Add to $W$, exactly one vertex, which is not in $S$ (as per Claim 5), from $V_{i,1,2}$, $V_{i,2,3}$, $V_{i,3,4}$, and $V_{i,1,4}$. It can
be verified that $W$ induces a $C_8$ in $G\oplus S$, which is a contradiction.
\end{proof}

Theorem~\ref{thm:ct} is a direct implication of Theorem~\ref{thm:pt}, Lemma~\ref{lem:inductive:cycle}, and Theorem~\ref{thm:c8}.
\begin{theorem}
\label{thm:ct}
Let $t\geq 8$ be any integer. Then \SCT\ $C_t$-free graphs is NP-complete.
Further, the problem cannot be solved in time $2^{o(|V(G)|)}$, unless the ETH fails. 
\end{theorem}



\bibliographystyle{plain}
\bibliography{main}
\end{document}